\documentclass{bioinfo}
\usepackage{hhline}
\usepackage{fancyhdr}
\usepackage[normalem]{ulem}
\usepackage{tabularx}
\usepackage{soul}
\usepackage{atbegshi,picture}
\usepackage{lipsum}
\usepackage{mathtools}
\usepackage{amsmath}
\usepackage{url}
\usepackage{comment}
\usepackage{setspace}

\renewcommand{\baselinestretch}{0.93} 

\usepackage{lipsum}                     
\usepackage{xargs}                      
\usepackage[pdftex,dvipsnames]{xcolor}  
\let\oldcitep=\citep 
\renewcommand{\citep}[1]{\textcolor[rgb]{0,0,1}{\oldcitep{#1}}}
\let\oldurl=\url 
\renewcommand{\url}[1]{\textcolor[rgb]{0,0,1}{\oldurl{#1}}}
\usepackage{floatrow}
\usepackage{graphicx}
\usepackage[caption=false]{subfig}

\usepackage{amssymb}
\usepackage{pifont}
\newcommand{\cmark}{\ding{51}}%
\newcommand{\xmark}{\ding{55}}%

\usepackage{tikz}

\usepackage{pdfpages}

\begin{document}
\doublespacing
\firstpage{1}

\subtitle{}

\title[COVIDHunter: An Accurate, Flexible, and Environment-Aware Open-Source COVID-19 Outbreak Simulation Model]{\center{\LARGE{COVIDHunter: An Accurate, Flexible, and Environment-Aware \\Open-Source COVID-19 Outbreak Simulation Model}}}

\author[Alser \textit{et~al}.]{\center{
Mohammed Alser, Jeremie S. Kim, Nour Almadhoun Alserr, \\Stefan W. Tell, and {Onur Mutlu}}}
\address{\center{
\large{ETH Zurich, Zurich 8006, Switzerland}
}}




\corresp{}
\history{}

\editor{}

\abstract{\textbf{Background:} 
{Early detection and isolation of COVID-19 patients are essential for successful implementation of mitigation strategies and eventually curbing the disease spread.
With a limited number of daily COVID-19 tests performed in every country, simulating the COVID-19 spread along with the potential effect of each mitigation strategy currently remains one of the most effective ways in managing the healthcare system and guiding policy-makers. \\
\textbf{Methods:} 
We introduce \emph{COVIDHunter}, a flexible and accurate COVID-19 outbreak simulation model that evaluates the current mitigation measures that are applied to a region and provides suggestions on what strength the upcoming mitigation measure should be.
The key idea of COVIDHunter is to quantify the spread of COVID-19 in a geographical region by simulating the average number of new infections caused by an infected person considering the effect of external factors, such as environmental conditions (e.g., climate{, temperature, humidity}) and mitigation measures.
} \\
\textbf{Results:} 
{Using Switzerland as a {case study}, COVIDHunter estimates that if the {policy-makers 
relax the mitigation measures by 50\% for 30 days then both the \emph{daily} capacity need for hospital beds and \emph{daily} number of deaths increase exponentially by an average of $5.1\times$, who may occupy ICU beds and ventilators for a period of time}.
Unlike existing models, the COVIDHunter model accurately monitors and predicts the daily number of cases, hospitalizations, and deaths due to COVID-19. 
Our model is flexible to configure and simple to modify for modeling different scenarios under different environmental conditions and mitigation measures.} \\
\textbf{Availability:} We release the source code of the COVIDHunter implementation at https://github.com/CMU-SAFARI/COVIDHunter and show how to flexibly configure our model for any scenario and easily extend it for different measures and conditions than we account for. \\
\\}
\maketitle

\section{Introduction} \label{sec:introduction}

\emph{Coronavirus disease 2019} (COVID-19) is caused by SARS-CoV-2 virus, which was first detected in Wuhan, the capital city of Hubei Province in China, in early December 2019~\citep{du2020outbreak}. Since then, it has rapidly spread to nearly every corner of the globe and has been declared a pandemic in March 2020 by the World Health Organization (WHO). As of January 2021, COVID‑19 has since resulted in more than 96 million laboratory-confirmed cases around the world, and has killed nearly 2.2\% of the infected population. 
As there are currently no anti-SARS-CoV-2-specific drugs or effective vaccines widely available to everyone, early detection and isolation of COVID-19 patients remain essential for effectively curbing the disease spread. 
As a result, many countries across the world have implemented unprecedented lockdown and social distancing measures, affecting millions of people. Regardless of the availability and affordability of COVID-19 testing, it is still extremely challenging to detect and isolate COVID-19 infections at early stages due to three key issues. 
1) It is very difficult to accurately identify the initial contraction time of COVID-19 for a patient. This is because COVID-19 patients can develop symptoms between 2 to 14 days (or longer in a few cases) after exposure to the new coronavirus~\citep{lauer2020incubation, li2020early}. This variable delay is referred to as the virus' \emph{incubation period}.
2) The coronavirus genome can exhibit rapid genetic changes in its nucleotide sequence, which may occur during viral cell replication, within the host body, or during transmission between hosts~\citep{andersen2020proximal}. This genetic diversity affects the virus virulence, infectivity, transmissibility, and evasion of the host immune responses~\citep{phan2020genetic, pachetti2020emerging, toyoshima2020sars}.
3) The situation becomes even worse as the coronavirus can survive and therefore remain \emph{infectious} outside the host, on common surfaces such as metal, glass, and banknotes (both paper and polymer) at room temperature for \emph{up to 28 days}~\citep{kampf2020persistence, riddell2020effect}.

Simulating the spread of COVID-19 has the potential to mitigate the effects of the three key issues, help to better manage the healthcare system, and provide guidance to policy-makers on the effectiveness of various (current, planned or discussed) social distancing and mitigation measures. 
To this end, many COVID-19 simulation models are proposed (e.g., \citep{tradigo2020method, russell2020reconstructing, ashcroft2020covid}), some of which are announced to \emph{assist} in decision-making for policy-makers in countries such as the United Kingdom (ICL~\citep{flaxman2020estimating}), United States (IHME~\citep{reiner2020modeling}), and Switzerland (IBZ~\citep{huisman2020estimation}).
These models tend to follow one of two key approaches. (1) Evaluating the current actual epidemiological situation by accounting for reporting delays and under-reporting due to inefficiencies such as low number of COVID-19 tests. (2) Evaluating the current and future epidemiological situation by simulating the COVID-19 outbreak \emph{without} relying on the observed (laboratory-confirmed) number of cases in simulation.

The first approach, taken by the IBZ~\citep{huisman2020estimation}, LSHTM~\citep{russell2020reconstructing}, and \citep{ashcroft2020covid} models, is \emph{not} mainly used for prediction purposes as it reflects the epidemiological situation with about two weeks of time delay (due to its dependence on observed COVID-19 reports).
The IBZ model~\citep{huisman2020estimation} estimates the daily reproduction number, $R$, of SARS-CoV-2 from observed COVID-19 incidence time series data after accounting for reporting delays and under-reporting using the numbers of confirmed hospitalizations and deaths. 
The $R$ number describes how a pathogen spreads in a particular population by quantifying the average number of new infections caused by each infected person at a given point in time.
The LSHTM model~\citep{russell2020reconstructing} adjusts the daily number of observed COVID-19 cases by accounting for under-reporting (uncertainty) using both deaths-to-cases ratio estimates and correcting for delays between case confirmation (i.e., laboratory-confirmed infection) to death.

The second approach, taken by ICL~\citep{flaxman2020estimating} and IHME~\citep{reiner2020modeling} models, usually requires a large number of various input parameters and assumptions. 
IHME~\citep{reiner2020modeling} model requires input parameters such as testing rates, mobility, social distancing policies, population density, altitude, smoking rates, self-reported contacts, and mask use. This model makes two key assumptions: 1) the infection fatality rate (IFR), which indicates the rate of people that die from the infection is taken using data from the Diamond Princess Cruise ship and New Zealand and 2) the decreasing fatality rate is reflective of increased testing rates (identifying higher rates of asymptomatic cases).
ICL~\citep{flaxman2020estimating} model requires input parameters such as the daily number of confirmed deaths, IFR, mobility rates from Google, age- and country-specific data on demographics, patterns of social contact, and hospital availability. This model makes three key assumptions: 1) age-specific IFRs observed in China and Europe are the same across every country, 2) the number of confirmed deaths is equal to the true number of COVID-19 deaths, and 3) the change in transmission rates is a function of average mobility trends. 


\begin{table*}[hbt!]
\footnotesize
\begin{center}
\begin{tabular}{ rccccccc }
\hline 
    & \multicolumn{1}{c}{\textbf{Open}} & \textbf{Well-} & \textbf{Accounting for} & \textbf{Low Number} & \textbf{Reported} & \\ 
      \multicolumn{1}{c}{\textbf{Model}}
      
	& \textbf{Source} & \textbf{Documented$^\#$} & \textbf{Weather Changes} & \textbf{of Parameters} &\textbf{COVID-19 Statistics} &  \\
\hline
COVIDHunter (this work)
	& \cmark
	& \cmark
	& \cmark
    & \cmark
    & \cmark~($R$, cases, hospitalizations, and deaths) \\
IBZ~\citep{huisman2020estimation}
& \cmark & \xmark & \xmark & \cmark & \xmark~(only $R$) & \\ LSHTM~\citep{russell2020reconstructing} & \cmark & \xmark & \xmark & \cmark & \xmark~(only cases) & \\ 
ICL~\citep{flaxman2020estimating}
& \cmark & \cmark & \xmark & \xmark & \cmark~($R$, cases, hospitalizations, and deaths) & \\ 
IHME~\citep{reiner2020modeling}
& \cmark$^*$ & \xmark & \xmark & \xmark & \xmark~(cases, hospitalizations, and deaths) & \\ 

\hline
\end{tabular}
\begin{flushleft}\footnotesize{$^\#$ Based on each model's GitHub page {(all models are available on GitHub)}. \hspace{0.1cm} $^*$ The available packages are configured \emph{only} for the IHME infrastructure.}\end{flushleft}
\caption{Comparison to other models used to inform government policymakers, as of January 2021.} 
\label{tab:prior_works}
\end{center}
\vspace{-20 pt}
\end{table*}

To our knowledge, there is currently no model capable of accurately monitoring the current epidemiological situation and predicting future scenarios while considering a reasonably low number of parameters and accounting for the effects of environmental conditions, as we summarize in Table~\ref{tab:prior_works}. 
The low number of parameters provides four key advantages: 1) allowing flexible (easy-to-adjust) configuration of the model input parameters for different scenarios and different geographical regions, 
2) enabling short simulation execution time and simpler modeling, 
3) enabling easy validation/correction of the model prediction outcomes by adjusting fewer variables,
and 4) being extremely useful and powerful especially during the early stages of a pandemic as many of the parameters are unknown.
Simulation models need to consider the fact that 
the environmental conditions (e.g., {air} temperature)
affect pathogen infectivity~\citep{fares2013factors, kampf2020persistence, riddell2020effect, xu2020modest} and simulating this effect helps to provide accurate estimation of the epidemiological situation.

Our \textbf{goal} in this work is to develop such a COVID-19 outbreak simulation model.
To this end, we introduce \emph{COVIDHunter}, a simulation model that evaluates the current mitigation measures (i.e., non-pharmaceutical intervention or NPI) that are applied to a region and provides insight into what strength the upcoming mitigation measure should be and for how long it should be applied, while considering the potential effect of environmental conditions.
Our model accurately forecasts the numbers of infected and hospitalized patients, and deaths for a given day, as validated on historical COVID-19 data (after accounting for under-reporting).
The \textbf{key idea} of COVIDHunter is to quantify the spread of COVID-19 in a geographical region by calculating the daily reproduction number, $R$, of COVID-19 and scaling the reproduction number based on changes in both mitigation measures and environmental conditions.
The $R$ number changes during the course of the pandemic due to the change in the ability of a pathogen to establish an infection during a season and mitigation measures that lead to {lower number of} of susceptible individuals.
COVIDHunter simulates the entire population of a region and assigns \emph{each} individual in the population to a stage of the COVID-19 infection (e.g., from being healthy to being short-term immune to COVID-19) based on the scaled $R$ number.
Our model is flexible to configure and simple to modify for modeling different scenarios as it uses \emph{only} three input parameters, two of which are time-varying parameters, to calculate the $R$ number.
Whenever {applicable}, we compare the simulation output of our model to that of four state-of-the-art models currently used to inform policy-makers, IBZ~\citep{huisman2020estimation}, LSHTM~\citep{russell2020reconstructing}, ICL~\citep{flaxman2020estimating}, and IHME~\citep{reiner2020modeling}.

The \textbf{contributions} of this paper are as follows: 
\begin{itemize}
\item We introduce COVIDHunter, a flexible {and} validated simulation model that evaluates the current and future epidemiological situation by simulating the COVID-19 outbreak.
COVIDHunter accurately forecasts for a given day 1) the reproduction number, 2) the number of infected people, 3) the number of hospitalized people, 4) the number of deaths, and 5) number of individuals at each stage of the COVID-19 infection.
COVIDHunter evaluates the effect of different current and future mitigation measures on the COVIDHunter's five numbers.

\item As a case study, we statistically analyze the relationship between temperature and number of COVID-19 cases in Switzerland.
We find that for each 1$^{\circ}$C rise in daytime temperature, there is a 3.67\% decrease in the daily number of confirmed cases. We demonstrate how \emph{considering} the effect of {climate {(e.g., daytime temperature)} on} COVID-19 spread {significantly} improves the {prediction} accuracy.

\item {Compared to IBZ, LSHTM, ICL, and IHME models, COVIDHunter achieves more accurate estimation, provides no prediction delay, 
and provides ease of use and high flexibility due to the simple modeling approach that uses a small number of parameters}. 

\item {Using COVIDHunter, we demonstrate that the spread of COVID-19 in Switzerland is still active (i.e., $R$ > 1.0) and curbing this spread requires {maintaining the same or greater strength} of the currently applied mitigation measures for at least another 30 days}.




\item We release {the well-documented} source code of COVIDHunter and show {how easy it is} to flexibly configure for \emph{any} scenario and extend for different measures and conditions than we account for.
\end{itemize}
\vspace{-20 pt}
\section{Methods} \label{sec:methods1}

\subsection{Overview}
The primary purpose of our COVIDHunter model is to \emph{monitor and predict} the spread of COVID-19 in a flexibly-configurable and easy-to-use way, while accounting for changes in mitigation measures and environmental conditions over time.
We employ a three-stage approach to develop and deploy this model. 
(1) The COVIDHunter model predicts the daily $R$ value based on \emph{only} three input parameters to maintain both quick simulation and high flexibility in configuring these parameters. 
Each input parameter is configured based on either existing research findings or user-defined values.
Our model allows for \emph{directly} leveraging existing models that study the effect of \emph{only} mitigation measures (or \emph{only} environmental conditions) on the spread of COVID-19, as we show in Section~\ref{PacmanModel}.
(2) The COVIDHunter model predicts the number of COVID-19 cases based on the predicted $R$ number.
COVIDHunter simulates the entire population of a region and labels each individual according to different stages of the COVID-19 infection timeline.
Each stage has a different degree of infectiousness and contagiousness.
The model simulates these stages for each individual to maintain accurate predictions.
(3) The COVIDHunter model predicts the number of hospitalizations and deaths based on both the predicted number of cases and the $R$ number. Next, we explain the COVIDHunter model in detail.

\subsection{How does the COVIDHunter Model Work?}
\label{PacmanModel}
The COVIDHunter model predicts the dynamic value of $R$ for a population at a given day while considering three key factors: 1) the transmissibility of an infection into a susceptible host population, 2) mitigation measures (e.g., lockdown, social distancing, and isolating infected people), and 3) environmental conditions (e.g., {air} temperature).
Our model calculates the {time-varying} $R$ number using Equation \ref{EQU1} as follows:
\begin{equation}
\label{EQU1}
R(t) = R0 * (1-M(t)) * C_{e}(t)
\end{equation}
\noindent The $R$ number for a given day, $t$, is calculated by multiplying three terms: 1) the base reproduction number ($R0$) for the subject virus, 2) one minus the mitigation coeﬃcient ($M$), for the given day $t$ and 3) the environmental coeﬃcient ($C_{e}$) for the given day $t$.

The $R0$ number quantifies the transmissibility of an infection into a susceptible host population by calculating the expected average number of new infections caused by an infected person in a population with no prior immunity to a specific virus (as a pandemic virus is by definition novel to all populations). 
Hence, the $R0$ number represents the transmissibility of an infection at only the beginning of the outbreak assuming the population is not protected via vaccination.
Unlike the $R$ number, $R0$ number is a fixed value and it does not depend on time.
The $R$ number is a time-dependent variable that accounts for the population's reduced susceptibility.
The $R0$ number for the COVID-19 virus can be obtained from several existing studies (such as in~\citep{anastassopoulou2020data, Hilton2020Rnumber, chang2020modelling, shi2020effective, de2020epidemiological, rahman2020basic}) that estimate it by modeling contact patterns during the first wave of the pandemic.

The mitigation coeﬃcient ($M$) applied to the population is a time-dependent variable and it has a value between 0 and 1, where 1 represents the strongest mitigation measure and 0 represents no mitigation {measure} applied.
In different countries, mitigation {measures} take different forms, such as social distancing, self-isolation, school closure, banning public events, and complete lockdown.
These measures exhibit significant heterogeneity and differ in timing and intensity across countries~\citep{hale2020variation, davies2020effects}. 
Quantifying the mitigation measures on a scale from 0 to 1 across different countries is challenging.
\emph{The Oxford Stringency Index}~\citep{hale2020variation} maintains a twice-weekly-updated index that takes values from 0 to 100, representing the severity of nine mitigation measures that are applied by more than 160 countries. 
Another study~\citep{brauner2020inferring} estimates the effect of \emph{only} seven mitigation measures on the $R$ number in 41 countries.
We can \emph{directly} leverage such studies for calculating the mitigation coeﬃcient on a given day after changing the scale from 0:100 to 0:1 by dividing each value of, for example, the Oxford Stringency Index {by} 100. 

The environmental coeﬃcient ($C_{e}$) is a time-dependent variable representing the effect of external environmental factors on the spread of COVID-19 and it has a value 
between 0 and 2.
Several related viral infections, such as {the} Influenza virus, human coronavirus, and human respiratory, already show notable seasonality (showing peak incidences during \emph{only} the winter (or summer) months)~\citep{moriyama2020seasonality, fisman2012seasonality}. 
The seasonal changes in temperature, humidity, and ultraviolet light affect the pathogen {infectiousness} outside the host~\citep{fares2013factors, kampf2020persistence, riddell2020effect, xu2020modest}.
However, the indoor environmental conditions are usually well-controlled throughout the year, where human behavior and number of households can be the major contributor to the spread of the COVID-19~\citep{moriyama2020seasonality}.
There are currently several studies that demonstrate the strong dependence of the transmission of SARS-CoV-2 virus on one or more environmental conditions, {\emph{even after}} controlling (isolating) the impact of mitigation measures and behavioral changes that reduce contacts.
Several studies have demonstrated increased infectiousness by {a} country-dependent fixed-rate with each 1$^{\circ}$C fall in daytime temperature~\citep{xie2020association, prata2020temperature}.
Another study supports the same temperature-infectiousness relationship, but it also finds that before applying any mitigation measures, a one degree drop in relative humidity shows increased infectiousness by a rate lower (2.94$\times$ less) than that of temperature~\citep{wang2020high}.
{Another study follows a simple way of modeling the effect of seasonality on COVID-19 transmission using a sinusoidal function with an annual period}~\citep{noll2020covid}.

One of the most comprehensive studies that spans more than 3700 locations around the world is \emph{HARVARD CRW}~\citep{xu2020modest}.
It finds the statistical correlation between the relative changes in the $R$ number and both weather (temperature, ultraviolet index, humidity, air pressure, and precipitation) and air pollution (SO2 and Ozone) after controlling the impact of mitigation measures.
The study provides a \emph{CRW Index} that has a value from 0.5 to 1.5.
The  {percentage} difference between any two consecutive values provided by the CRW Index represents the effect that both weather and air pollutants have on the $R$ number.
For example, a drop in the CRW Index by 10\% in a given location points to a 10\% reduction in the $R$ number due to weather changes and air pollutants.
Our model enables applying  {\emph{any of these studies}} by adjusting our environmental coeﬃcient on a given day, as we experimentally demonstrate in Section~\ref{sec:results}.
For example, if the {COVIDHunter} user chooses to consider the HARVARD CRW study, and the CRW Index shows, for example, a 10\% drop compared to its immediately preceding data point, then the environmental coeﬃcient of COVIDHunter should be 0.9 so that the $R$ value decreases by also 10\%.
Next, we explain how our model forecasts the number of COVID-19 cases based on Equation~\ref{EQU1}.

\subsection{Predicting the Number of COVID-19 Cases}

COVIDHunter tracks the number of infected and uninfected persons over time by clustering the population into four main categories: \texttt{HEALTHY}, \texttt{INFECTED}, \texttt{CONTAGIOUS}, and \texttt{IMMUNE}.
The model initially considers the entire population as uninfected (i.e., \texttt{HEALTHY}). 
For each simulated day, the model calculates the $R$ value using Equation~\ref{EQU1} and decides how many persons can be infected during that day.
The day when the first case of infection in a population introduced is defined by the user.
For each newly infected person (\texttt{INFECTED}), the model maintains a counter that counts the number of days from being infected to being contagious (\texttt{CONTAGIOUS}).
Several COVID-19 case studies show that \emph{presymptomatic} transmission can occur 1–3 days before symptom onset~\citep{wei2020presymptomatic, slifka2020presymptomatic}. 
COVID-19 patients can develop symptoms mostly after an incubation period of 1 to 14 days (the median incubation period is estimated to be 4.5 to 5.8 days)~\citep{lauer2020incubation, li2020early}.
We calculate the number of days of being contagious after being infected as a random number with a Gaussian distribution that has user-defined lowest and highest values.
Each contagious person may infect $N$ other persons depending on mobility, population density, number of households, and several other factors~\citep{ferguson2020report}.
We calculate the value of $N$ to be 
a random number with a Gaussian distribution that has the lowest value of 0 and the highest value determined by the user.
If $N$ is greater than the $R$ number (i.e., the target number of infections for that day has been reached){,}
further infections are curtailed preventing overestimation of $N$ by infecting \emph{only} $R$ persons.
Once the contagious person infects the desired number of susceptible persons, the status of the contagious person becomes immune (\texttt{IMMUNE}).
The immune status indicates that the person has immunity to reinfection due to either vaccination or being recently infected~\citep{lumley2020antibody, jagannathan2021immunity}. 

Our model also simulates the effect of infected travelers (e.g., daily cross-border commuters within the European Union) on the value of $R$. 
These travelers can initiate the infection(s) at the beginning of the pandemic.
If such infected travelers are absent (due to, for example, emergency lockdown) from the target population, the virus would die out once the value of $R$ decreases below 1 for a sufficient period of time. 
Both the number and percentage of infected travelers entering a region are configurable in our model. The percentage of incoming infected travelers is \emph{not} affected by the changes in the local mitigation measures, as these travelers were infected abroad.

Our model predicts the \emph{daily} number of COVID-19 cases for a given day {$t$}, as follows:

\begin{equation}
\label{EQU2}
Daily\_Cases(t) = \sum_{n=0}^{T_{INF}(t)} N(n) + \sum_{m=0}^{U_{CON}(t)} N(m)
\end{equation}

\noindent where $T_{INF}$ is the daily number of infected travelers that is a user-defined variable, $N()$ is a function that calculates the number of persons to be infected by a given person as a random number with a Gaussian distribution, and $U_{CON}$ is the daily number of contagious persons calculated by our model.

\subsection{Predicting the Number of COVID-19 Hospitalizations and Deaths}
\label{sec:HospDeath}

There are currently two key approaches for calculating the estimated number of both hospitalizations and deaths due to COVID-19: 1) using historical statistical probabilities, each of which is unique to each age group in a population~\citep{bhatia2020estimating, bi2020epidemiology} and 2) using historical COVID-19 hospitalizations-to-cases and deaths-to-cases ratios~\citep{kobayashi2020communicating}.
We choose to follow a modified version of the second approach as it does {\emph{not}} require 1) clustering the population into age-groups and 2) calculating the risk of each individual using the given probability, which both affect the complexity of the model and the simulation time. 

The number of COVID-19 hospitalizations for a given day, $t$, can be calculated as follows:
\begin{equation}
\label{EQU3}
Daily\_Hospitalizations(t) = Daily\_Cases(t) * X * C_{X}
\end{equation}

\noindent where $Daily\_Cases(t)$ is calculated using Equation~\ref{EQU2} and $X$ is the hospitalizations-to-cases ratio that is calculated as the average of daily ratios of the number of COVID-19 hospitalizations to the laboratory-confirmed number of COVID-19 cases.
As the {\emph{true}} number of cases is unknown due to lack of population-scale testing, it is extremely difficult to make accurate estimates of the {\emph{true}} number of COVID-19 hospitalizations~\citep{petropoulos2020forecasting}.
{As such}, we assume a fixed multiplicative relationship between the number of laboratory-confirmed cases and the {\emph{true}} number of cases.
We use the user-defined correction coefficient, $C_{X}$, of the hospitalizations-to-cases ratio to account for such a multiplicative relationship.

The number of COVID-19 deaths for a given day $t$ can be calculated as follows:
\begin{equation}
\label{EQU4}
Daily\_Deaths(t) = Daily\_Cases(t) * Y * C_{Y}
\end{equation}

\noindent where $Daily\_Cases(t)$ is calculated using Equation~\ref{EQU2} and $Y$ is the deaths-to-cases ratio, which is calculated as the average of daily ratios of the number of COVID-19 deaths to the number of COVID-19 laboratory-confirmed cases.
The {observed} number of COVID-19 deaths can still be less than the {\emph{true}} number of COVID-19 deaths due to, for example, under-reporting.
We use the user-defined correction coefficient, $C_{Y}$, to account for the under-reporting.
One way to find the \emph{true} number of COVID-19 deaths is to calculate the number of excess deaths.
The number of excess deaths is the difference between the observed {number} of deaths during time period and expected (based on historical data) {number} of deaths during the same time period.
For this reason, $C_{Y}$ may not necessarily be equal to $C_{X}$.



\subsection{Model Validation} \label{sec:modelvalidation}

We can validate our model using two key approaches.
1) Comparing the daily $R$ number predicted by our model (using Equation~\ref{EQU1}) with the daily {reported  official} $R$ number for the same region.
2) Comparing the daily number of COVID-19 cases predicted by our model (using Equation~\ref{EQU2}) with the daily number of laboratory-confirmed COVID-19 cases.
As of 2021, we {have} already witnessed more than one year of the pandemic, which provides us several observations and lessons.
The most obvious source of uncertainty, affecting \emph{all} models, is that the {\emph{true}} number of persons that are previously infected or currently infected is \emph{unknown}~\citep{wilke2020predicting}.
This affects the accuracy of the reported $R$ number since it is calculated as, for example, the ratio of the number of cases for a week (7-day rolling average) to the number of cases for the preceding week.
Adjusting the parameters of our model to fit the curve of the number of confirmed cases is likely to be highly uncertain.
The publicly-available number of COVID-19 hospitalizations and deaths can {provide} more reliable data.

For these reasons, we decide to use a combination of reported numbers of cases, hospitalizations, and deaths for validating our model using three key steps.
1) We leverage the more reliable data of reported number of hospitalizations (or deaths) to estimate the {\emph{true}} number of COVID-19 cases using the ratio of number of laboratory-confirmed hospitalizations (or deaths) to the number of laboratory-confirmed cases during the second wave of the COVID-19 pandemic.
We assume that the COVID-19 statistics during the second wave is more accurate than {that} during the first wave {because} generally more testing is performed {in the second wave}.
2) We consider a multiplicative relationship {between} the {\emph{true}} number of COVID-19 cases and that estimated in step 1.
In our experimental evaluation (Section~\ref{sec:results}), we use the {\emph{true}} number of COVID-19 cases calculated using different multiplicative factor values (we refer to them as \emph{certainty rate levels}) as a ground-truth for validating our model.
A certainty rate of, for example, 50\% means that the {\emph{true}} number of COVID-19 cases is actually \emph{double} that calculated in step 1.
3) We use our model to calculate both the daily $R$ number (using Equation~\ref{EQU1}) and the number of COVID-19 cases (using Equation~\ref{EQU2}).
We fix the two terms of Equation~\ref{EQU1}, $R0$ and $C_{e}$, using publicly-available data for a given region and change the third term, $M$, until we fit the curve of the number of cases predicted by our model to the ground-truth plot calculated in step 2. 
We use the same methodology to validate our predicted numbers of hospitalizations and deaths with different certainty rate levels as we show in Section~\ref{sec:results} and \textcolor{blue}{Supplementary Excel File (SimulationResultsForSwitzerland.xlsx)}. 


\subsection{Flexibility and Extensibility of {the} COVIDHunter Model}

We especially build COVIDHunter model to be flexible to configure and easy to extend for representing any existing or future scenario using different values of the three terms of Equation~\ref{EQU1}, 1) $R0$, 2) $M(t)$, 3) $C_{e}(t)$, in addition to several other parameters such as the population, number of travelers, percentage of expected infected travelers to the total number of travelers, and hospitalizations- or deaths-to-cases ratios.
Our modeling approach acts across the overall population without assuming any specific age structure for transmission dynamics. 
It is still \emph{possible} to consider each age group separately using individual runs of COVIDHunter model simulation, each of which has its own parameter values adjusted for the target age group.
{The} COVIDHunter model considers each location independently of other locations, but it also accounts for potential movement between locations by adjusting the corresponding parameters for travelers.
By allowing most of the parameters to vary in time, $t$, {the} COVIDHunter model is capable {of} accounting for any change in transmission intensity due to changes in environmental conditions and mitigation measures over time.
As we explain in Section~\ref{PacmanModel}, the flexibility of configuring the environmental coeﬃcient and mitigation coeﬃcient allows our proposed model to control for location-specific differences in population density, cultural practices, age distribution, and time-variant mitigation responses in each location.
Our modeling approach considers a single strain of the COVID-19 virus by using a single base reproduction number, $R0$.
It is possible to consider multiple virus strains by running the model simulation multiple times, each of which considers one of the strains individually.
The model can be extended to consider multiple virus strains by replacing the $R0$ number by multiple $R0$ numbers that represent the {different} strains~\citep{Reichmuth2021}.
\vspace{-10 pt}

\section{Results} \label{sec:results}
We evaluate the daily 1) $R$ number, 2) mitigation measures, and 3) numbers of COVID-19 cases, hospitalizations, deaths.
We also evaluate the daily numbers of \texttt{HEALTHY}, \texttt{INFECTED}, \texttt{CONTAGIOUS}, and \texttt{IMMUNE} in the \textcolor{blue}{Supplementary Excel File (SimulationResultsForSwitzerland.xlsx)}.
We compare the predicted values to their corresponding observed values whenever possible.
{We provide a comprehensive treatment of \emph{all} datasets, models, and evaluation results with different model configurations in the} \textcolor{blue}{Supplementary Materials}. 
We provide another prediction run in~\citep{alser2021covidhunterv1} for the period from 22 January 2021 until 22 February 2021, which was carried out on 22 January 2021.


\subsection{Determining {the} Value of Each Variable in the Equations}
We use Switzerland as a use-case for all the experiments.
However, our model is not limited to any specific region as the parameters it uses are completely configurable.
To predict the $R$ number, we use Equation~\ref{EQU1} that requires three key variables.
We set the base reproduction number, $R0$, for the SARS-CoV-2 in Switzerland as 2.7, as shown in~\citep{Hilton2020Rnumber, anastassopoulou2020data}.
We choose two main approaches for setting the {value} of the time-varying environmental coeﬃcient  variable {($C_{e}$)}.
1) Performing statistical analysis for the relationship between the daily number of COVID-19 cases and average daytime temperature in Switzerland.
As we provide in the~\textcolor{blue}{Supplementary Materials, Section 1}, our statistical analysis shows that each 1$^{\circ}$C rise in daytime temperature is associated with a 3.67\% ($t$-value = -3.244 and $p$-value = 0.0013) decrease in the daily number of {confirmed} COVID-19 cases.
We refer to this approach as Cases-Temperature Coefficient (CTC).
2) Applying the \emph{HARVARD CRW}~\citep{xu2020modest} (CRW in short), which provides the statistical relationship between the relative changes in the $R$ number and both weather factors and air pollutants after controlling {for} the impact of mitigation measures.
We change the daily mitigation coeﬃcient, $M(t)$, value based on the ratio of number of confirmed hospitalizations to the number of confirmed cases with two certainty rate {levels} of 100\% and 50\%, as we explain in detail in Section~\ref{sec:modelvalidation}.
This helps us to take into account uncertainty in the observed number of COVID-19 cases, hospitalizations, and deaths.
We set the minimum and maximum incubation time for SARS-CoV-2 as 1 and 5 days, respectively{, as 5-day period} represents the median incubation period worldwide~\citep{lauer2020incubation, li2020early}.
We set the population to 8654622.
We empirically choose the values of $N$, the number of travelers, and the ratio of the number of infected travelers to the total number of travelers to be 25, 100, 
and 15\%, respectively.

\subsection{Evaluating the Expected Number of COVID-19 Cases for Model Validation}
\label{results:modelvalidation}
\begin{figure}[H]
\centerline{\includegraphics[width=0.6\linewidth]{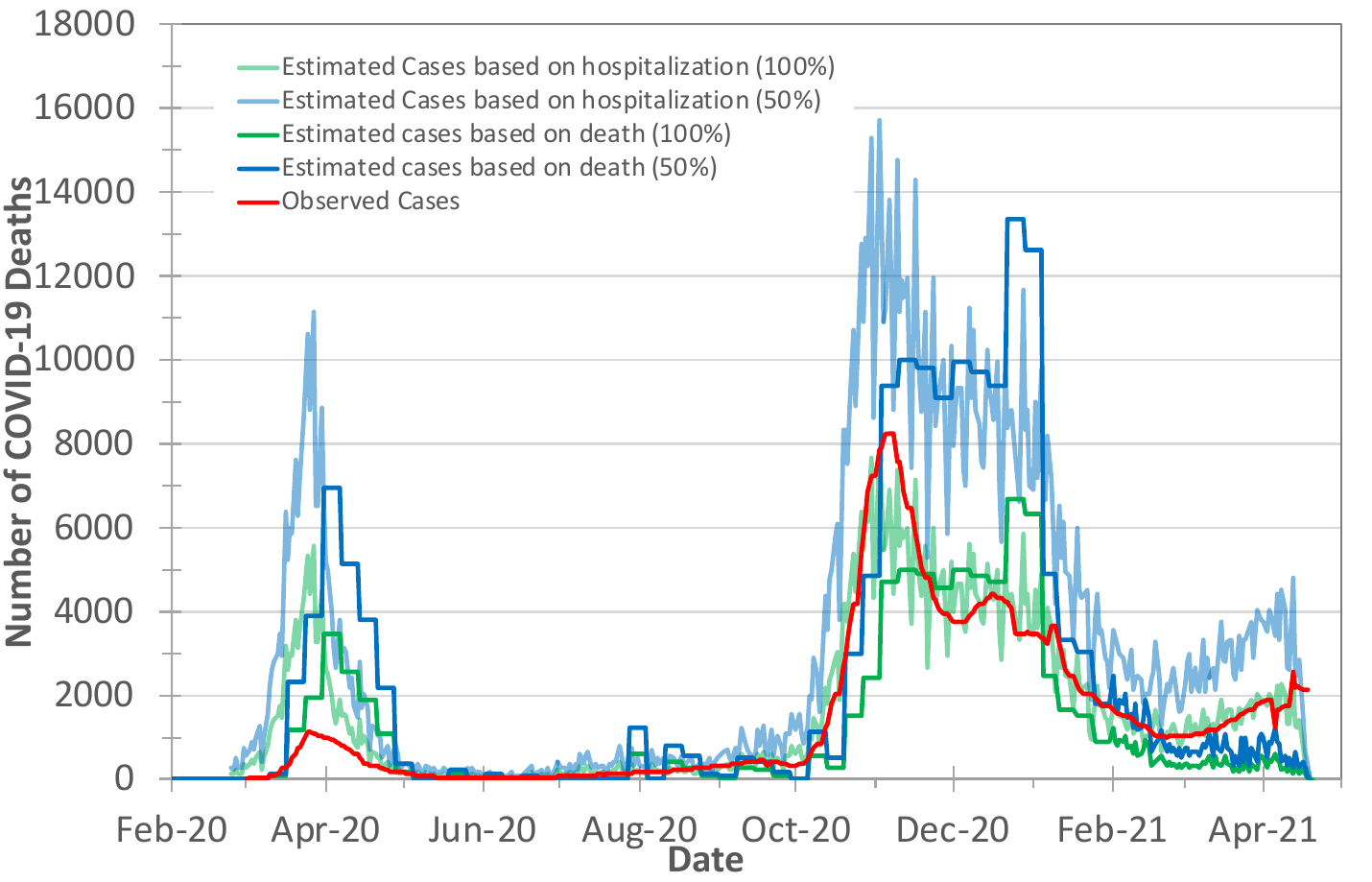}}
\caption{Observed {(officially reported)} and expected number of COVID-19 cases in Switzerland during the years of 2020 and 2021.
We calculate the expected number of cases based on both the hospitalizations-to-cases and deaths-to-cases ratios for the second wave. We assume two {certainty rate levels of}  50\% and 100\%.}
\label{FIG3}
\end{figure}

As the exact true number of COVID-19 cases remains unknown (due to, for example, lack of population-scale COVID-19 testing), we expect the true number of COVID-19 cases in Switzerland to be higher than the observed (laboratory-confirmed) number of cases.
We calculate the expected true number of cases based on both numbers of deaths and hospitalizations, as we explain in Section~\ref{sec:modelvalidation}. 
To account for possible missing number of COVID-19 deaths, we consider the {excess} deaths instead of observed deaths.
We calculate the {excess} deaths as the difference between 5-year average of weekly deaths and the observed weekly number of deaths in {both 2020 and 2021}. 
We find that $X$ (hospitalizations-to-cases ratio) and $Y$ (deaths-to-cases ratio, using {excess} {death} data) to be 3.75\% and 2.441\%, respectively, during the second wave of the pandemic in Switzerland.
We choose the second wave to calculate the values of $X$ and $Y$ as Switzerland {has increased} the daily number of COVID-19 testing by $5.31\times$ (21641/4074) on average compared to the first wave.
We calculate the expected number of cases {on} a given day $t$ with certainty rate levels of 100\% and 50\% based on hospitalizations by dividing the number of hospitalizations at $t$ by $X$ and $X/2$, respectively, as we show in Figure~\ref{FIG3}.
We apply the same approach to calculate the expected number of cases {on} a given day $t$ with certainty rate levels of 100\% and 50\% based on deaths using $Y$ and $Y/2$, respectively.

Based on Figure~\ref{FIG3}, we make two key observations.
1) The plot for the expected number of cases calculated based on the number of deaths is shifted forward by 10-20 days (15 days on average) from that for the expected number of cases calculated based on the number of hospitalizations.
This is due to the fact that each hospitalized patient usually spends {some} number of days in hospital before dying of COVID-19.
We do not observe a significant time shift between the plot of the expected number of cases calculated based on the number hospitalizations and the plot of observed (laboratory-confirmed) cases.
2) The expected number of cases calculated based on the number of hospitalizations is on {average $2.7\times$} higher than the expected number of cases calculated based on the number of deaths (after accounting for the 15-day shift) for the same certainty rate.
This is expected as not all hospitalized patients die.

We conclude that both numbers of hospitalizations and deaths can be used for estimating the true number of COVID-19 cases after accounting for the time-shift effect.



\subsection{Observed and Predicted $R$ number of SARS-CoV-2}
We calculate the predicted $R$ number using our model (Equation~\ref{EQU1}) and compare it to the observed {official} $R$ number and {the $R$ number} of two state-of-the-art models, ICL and IBZ, for the two years of 2020 and 2021. 
We configure COVIDHunter using the following configurations: 1) CTC as environmental condition approach, 2) certainty rate levels of 50\% and 100\%, and 3) mitigation coefficient {values of 0.35 and} 0.7.
{All our scripts are provided in our GitHub {page}.}
We consider the mean $R$ number provided by {the} ICL model.
We consider the median $R$ number calculated by {the} IBZ model based on observed number of hospitalized patients.
IBZ provides the predicted (after {9 April 2021}) $R$ number as the mean of the estimates from the last 7 days.

\begin{figure}[H]
\centerline{\includegraphics[width=0.6\linewidth]{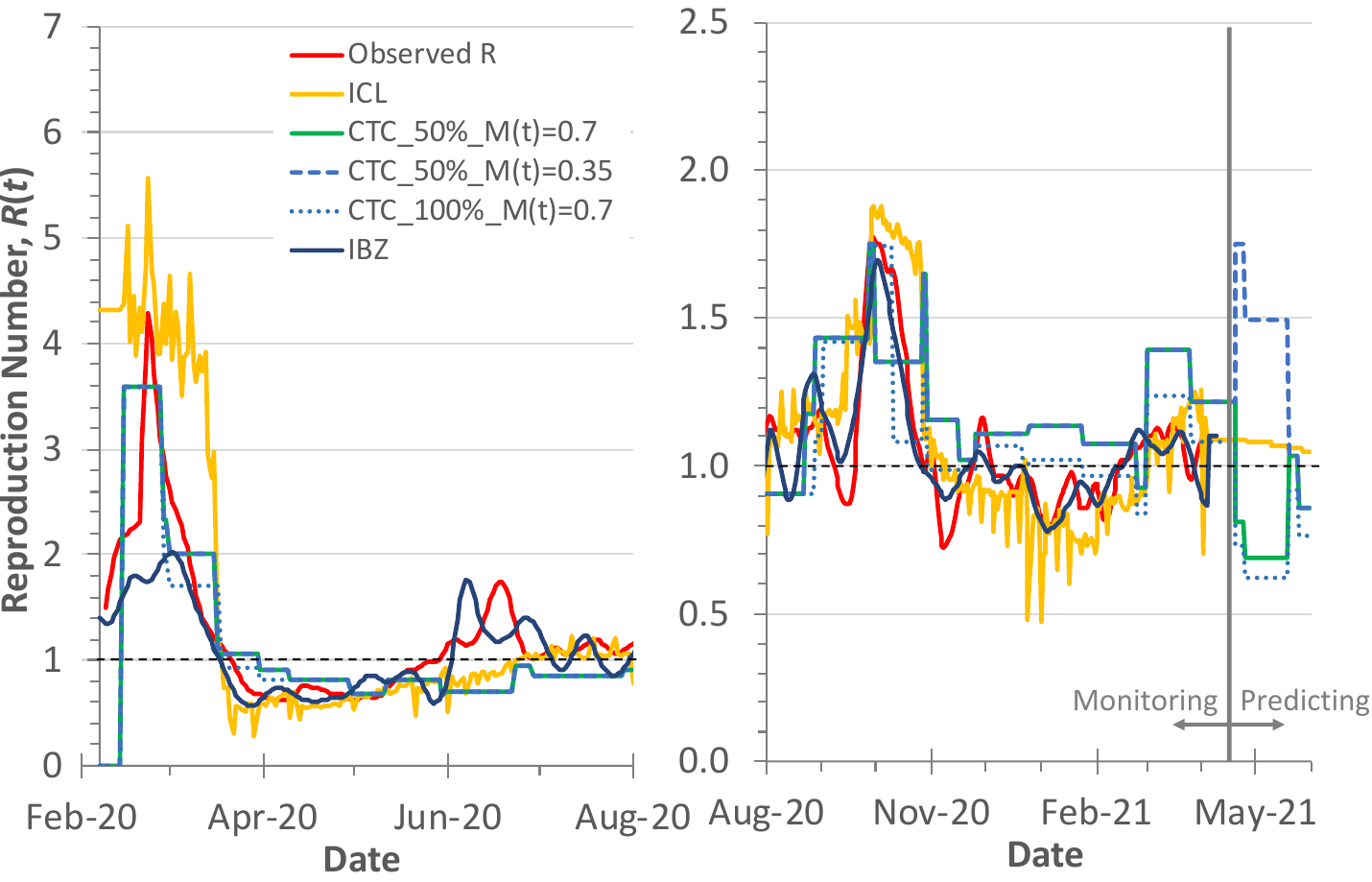}}
\caption{Observed and predicted reproduction number, $R(t)$, for the two years of 2020 and 2021. 
We use CTC environmental condition approach, {certainty rate levels of 50\% and} 100\%, and mitigation coefficient values of 0.35 and 0.7 for COVIDHunter.
We compare COVIDHunter's predicted $R$ number to the observed $R$ number and two state-of-the-art models, ICL and IBZ.
The horizontal dashed line represents $R(t)$ ={1.0}.
}
\label{FIG1}
\end{figure}

Based on Figure~\ref{FIG1}, we make three key observations.
1) COVIDHunter predicts the changes in $R$ number much (4-13 days) earlier than that predicted by ICL model, which leads to a more accurate prediction.
The $R$ number calculated by COVIDHunter (with a certainty rate level of 50\%) {before 19 April 2021 is on average 1.1$\times$ more} than that provided by ICL model, IBZ model, and the observed {official} $R$ number.
Using a certainty rate level of 100\%, COVIDHunter predicts the $R$ number to be close in value to the observed $R$ number.
{The $R$ numbers calculated by IBZ model and official authority (observed) are normally not provided for the last two weeks (as we discussed in the Section 1)}.
2) Our model predicts that the current $R$ number is still higher than 1 ({1.215 and 1.099 using certainty rate levels of 50\% and 100\%, respectively}) during {April 2021}.
This indicates that the spread of the SARS-CoV-2 virus is still active and it causes exponential increase in number of new cases.
3) 
{Our model predicts that if the mitigation measures that are applied nationwide in Switzerland are tightened (\emph{M(t)} increases from 0.55 to 0.7) for only 30 days (19 April to 19 May 2021), then the $R$ number decreases by at least 1.75$\times$ (from 1.215 to 0.691).
However, if the mitigation measures are relaxed (\emph{M(t)} drops from 0.55 to 0.35) for only 30 days (19 April to 19 May 2021), then the $R$ number increases by at least 1.23$\times$ (from 1.215 to 1.497)}.

{We conclude that COVIDHunter's estimation of the $R$ number is more accurate than that calculated by the ICL {and IBZ models}, as validated by the currently observed $R$ number}.

\subsection{Evaluating the Mitigation Measures}
\begin{figure}[H]
\centerline{\includegraphics[width=0.6\linewidth]{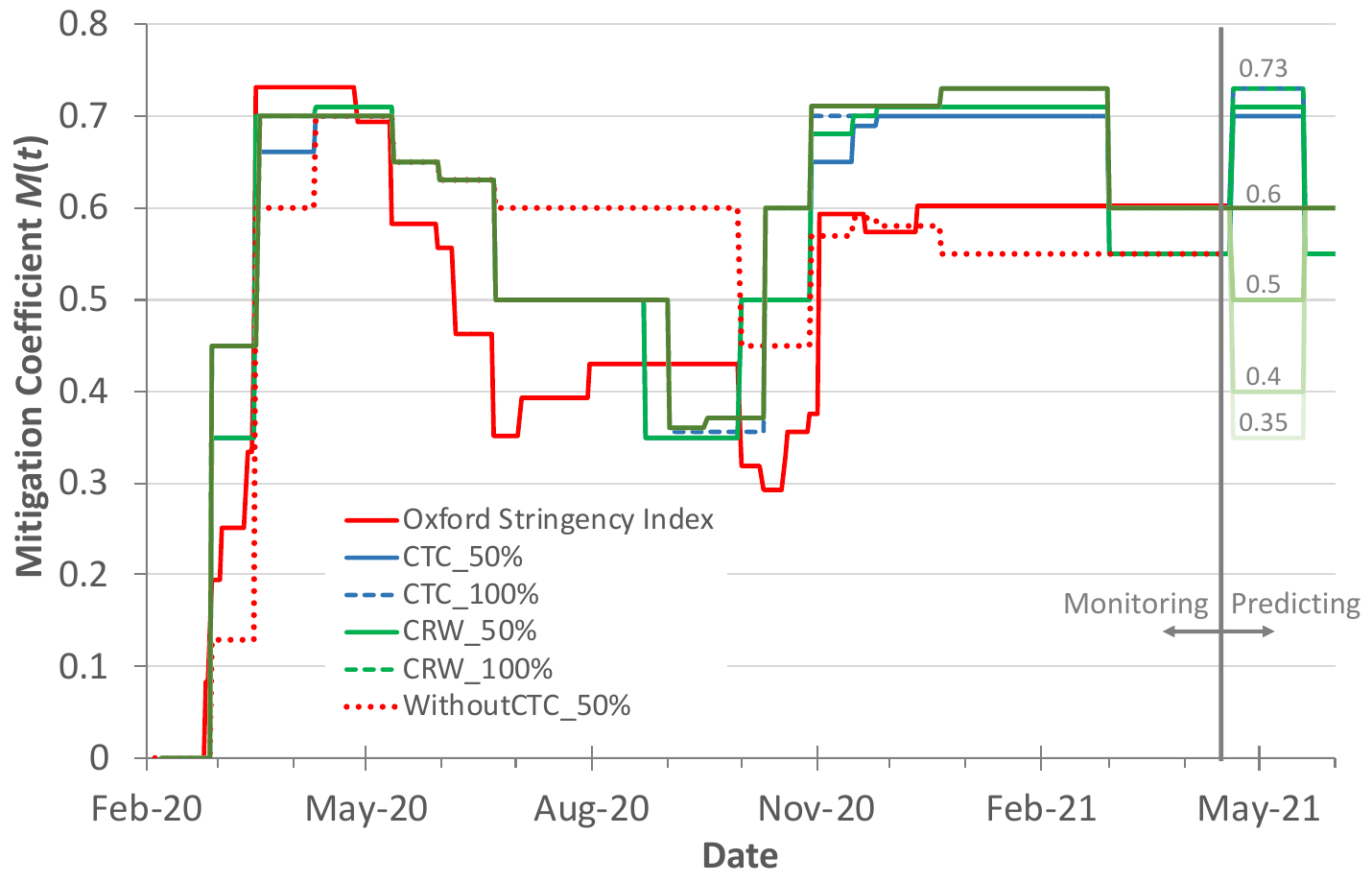}}
\caption{Predicted strength of the mitigation measures (mitigation coefficient, $M(t)$) applied in Switzerland from January 2020 to May 2021 provided by Oxford Stringency Index and COVIDHunter.
We use two different environmental condition approaches, CRW and CTC.
We assume two certainty rate {levels} of 50\% and 100\%. 
We use five mitigation $M(t)$ values of 0.35, 0.4, 0.5, 0.6, and 0.7 for each configuration of our model during {19 April to 19 May 2021}. 
The plot called \texttt{WithoutCTC\_50\%} represents the evaluation of the current mitigation measures while ignoring the effect of environmental changes.}
\label{FIG2}
\end{figure}

We evaluate the mitigation coefficient, $M(t)$, which represents the mitigation measures applied (or to be applied) in Switzerland from January 2020 to June 2021.
We use two different environmental condition approaches, CRW and CTC.
We assume two certainty {rate levels} of 50\% and 100\% to account for uncertainty in the observed number of cases.
We use five mitigation coefficients, $M(t)$, values of 0.35, 0.4, 0.5, 0.6, and 0.7 for each configuration of COVIDHunter during {19 April to 19 May 2021}.
We compare the evaluated mitigation measures to that evaluated by the Oxford Stringency Index~\citep{hale2020variation}, as we provide in Figure~\ref{FIG2}.
We also evaluate the mitigation coefficient when we ignore the effect of environmental changes (i.e., {by setting} $C_e$=1 in Equation~\ref{EQU1}), while maintaining the same number of COVID-19 cases of that provided with a certainty rate {level} of 50\%.

Based on Figure~\ref{FIG2}, we make four key observations.
1) {Excluding the effect of environmental changes from the COVIDHunter model, by setting $C_e$=1 in Equation}~\ref{EQU1}, {leads to an inaccurate evaluation of the mitigation measures.
For example, during the summer of 2020 (between the two major waves of 2020), COVIDHunter (\texttt{WithoutCTC\_50\%}) evaluates the mitigation coefficient to be as high as 0.6.
This means that the mitigation measures ({\emph{only}} mandatory of wearing mask on public transport) applied during the summer of 2020 are \emph{only} 14\% more relaxed compared to the mitigation measures (e.g., closure of schools, {restaurants}, and borders{, ban on small and large events}) applied during the first wave, which is {implausible}.} 
This highlights the importance of considering the effect of external environmental changes on simulating the spread of COVID-19.
Unfortunately, environmental change {effects} are \emph{not} considered by \emph{any} of {the} IBZ, LSHTM, ICL, and IHME models, which we believe {is} a serious shortcoming of these prior models.
2) A drop by 3-30\% (as we observe during the mid of November 2020 and the end of August 2020, respectively) in the strength of the mitigation measures for {a} certain period of time (10 to 20 days) is enough to double the predicted number of COVID-19 cases.
3) We evaluate the strength of the mitigation measures applied in Switzerland to be usually (65\% of the time) up to 80\% to 131\% higher than that provided by the Oxford Stringency Index.
4) The strength of the mitigation measures {has} changed 11 times and {2 times during the years of 2020 and 2021, respectively}, each of which is maintained for at least 9 days and at most 66 days (32 days on average). 

We conclude that considering the effect of environmental changes (e.g., daytime temperature) on the spread of COVID-19 improves simulation outcomes and provides accurate evaluation of the strength of the past and current mitigation measures.

\subsection{Evaluating the {Predicted N}umber of COVID-19 Cases}
\begin{figure}
\centerline{\includegraphics[width=0.6\linewidth]{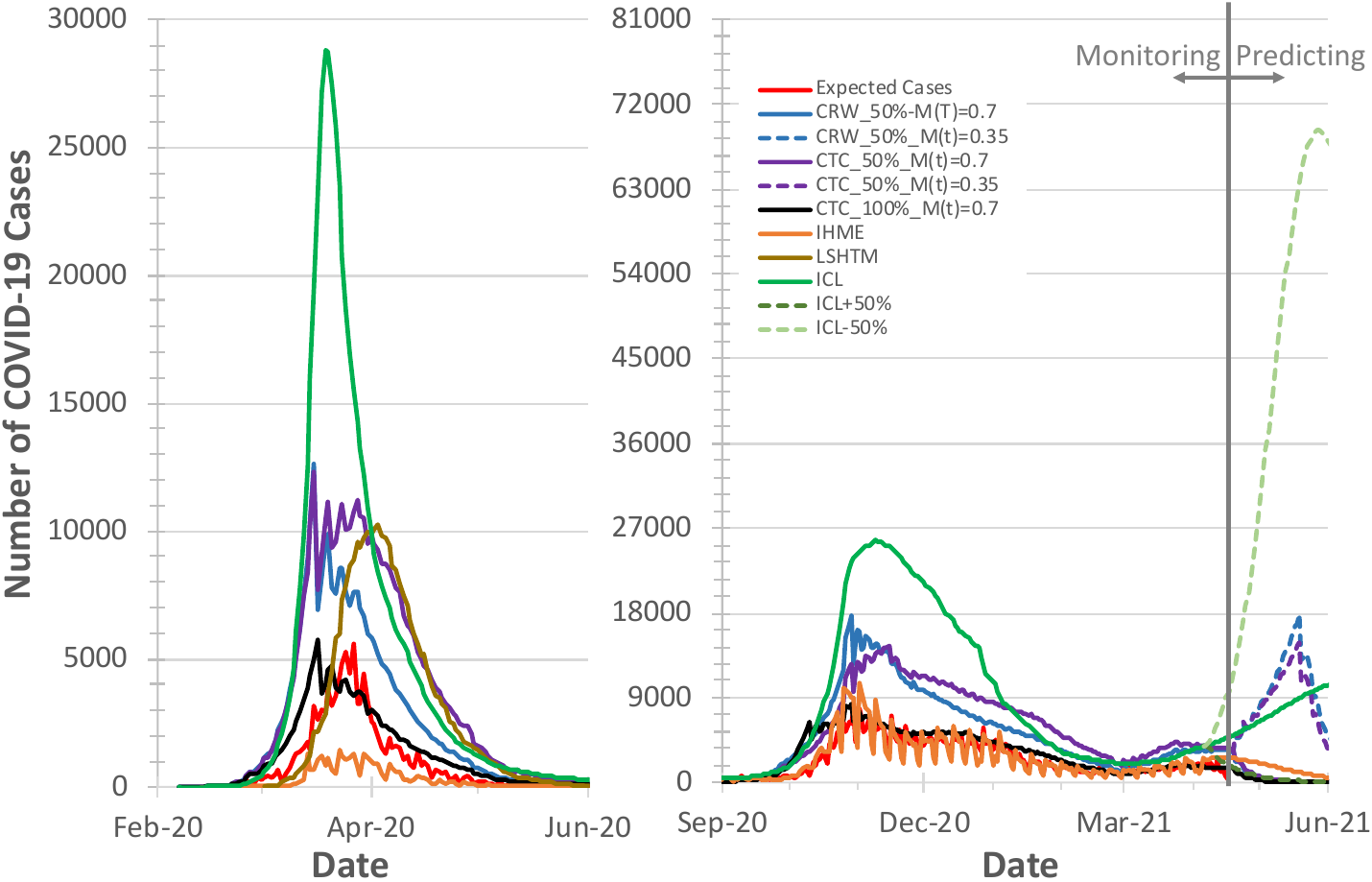}}
\caption{Observed and predicted number of COVID-19 cases by our model and other three state-of-the-art models.
We use two different environmental condition approaches, CRW and CTC with two certainty rate {levels} of 50\% and 100\%. 
We use {two} mitigation coefficient, $M(t)$, values of 0.35 and 0.7 for each configuration of our model during {19 April to 19 May 2021}.}
\label{FIG8}
\end{figure}

We evaluate COVIDHunter's {\emph{predicted}} daily number of COVID-19 cases in Switzerland.
We compare the predicted numbers by our model to the observed numbers and {those} provided by three state-of-the-art models (ICL, IHME, and LSHTM){,} as {shown} in Figure~\ref{FIG8}.
We calculate the observed number of cases as the expected number of cases with a certainty rate {level} of 100\% (as we discuss in Section~\ref{results:modelvalidation}).
We use three default configurations for the prediction of {the} ICL model: 1) {strengthening} mitigation measures by 50\%, 2) maintaining the same mitigation measures, and 3) relaxing mitigation measures by 50\%\, which we refer to as \texttt{ICL+50\%}, \texttt{ICL}, and \texttt{ICL-50\%}, respectively, in Figures~\ref{FIG8}, \ref{FIG5}, and \ref{FIG6}.
We use the mean numbers reported by the IHME model that represents the most relaxed mitigation measures, called as "no vaccine" by the IHME model.
We use the median numbers reported by {the} LSHTM model.

Based on Figure~\ref{FIG8}, we make four key observations.
1) Our model predicts that the number of COVID-19 cases reduces significantly (less than 50 daily cases) within May 2021 if the mitigation measures that are applied nationwide in Switzerland are tightened (\emph{M(t)} increases from 0.55 to 0.7) for at least 30 days. 
If the authority decides to relax the mitigation measures to the lowest strength that has been applied during the year of 2020 {(i.e., $M(t)=0.35$)}, then the daily expected number of cases increases by an average of {$5.1\times$ and $4.13\times$ (up to 17,892 daily cases)} using {the} CRW and CTC environmental approaches, respectively. {We provide a comprehensive evaluation for the effect of different mitigation coefficient values on the number of cases in the} \textcolor{blue}{Supplementary Materials, Section 2}. 
2) COVIDHunter (\texttt{CTC\_100\%\_M(t)=0.7}) predicts the number of COVID-19 cases to be equivalent to that predicted by {the} IHME model during the second wave with a certainty rate {level} of 100\%.
However, during the first wave, the prediction of {the} IHME model is {3.8~$\times$ less than the expected number of cases using a certainty rate level of 100\%}.
This means that, unlike our model, the IHME model considers the laboratory-confirmed cases during the first wave to be as if the tests are done at a population-scale, which is very likely incorrect. This is in line with a recent study~\citep{ioannidis2020forecasting} that demonstrates the high inaccuracy of the IHME model. 
3) Overall, our model {predicts} {up to $7.9\times$ and $6.4\times$ (on average $1.9\times$ and $2.1\times$)} {smaller} number of COVID-19 cases than that predicted by ICL model using CTC and CRW approaches, respectively, and a certainty rate of 50\%.
This suggests that the multiplicative relationship between the confirmed number of cases and the true number of cases can be represented by a certainty rate of 22\% to 33\%, which our model can easily account for.
4) The number of COVID-19 cases {estimated} by the LSHTM model during the first wave is 1) on average 24\% less than that {estimated} by COVIDHunter and 2) 10 days late from that predicted by COVIDHunter, IHME, and ICL.
The prediction of the LSHTM model during the second wave is not available by the model's pre-computed projections.

{We conclude that COVIDHunter provides more accurate estimation of the number of COVID-19 cases, compared to IHME (which provides inaccurate estimation during the first wave) and ICL (which provides over-estimation), with a complete control over the certainty rate level, mitigation measures, and environmental conditions. Unlike LSHTM, COVIDHunter also ensures no prediction delay}.

\subsection{Evaluating the {Predicted N}umber of COVID-19 Hospitalizations}
\begin{figure}
\centerline{\includegraphics[width=0.6\linewidth]{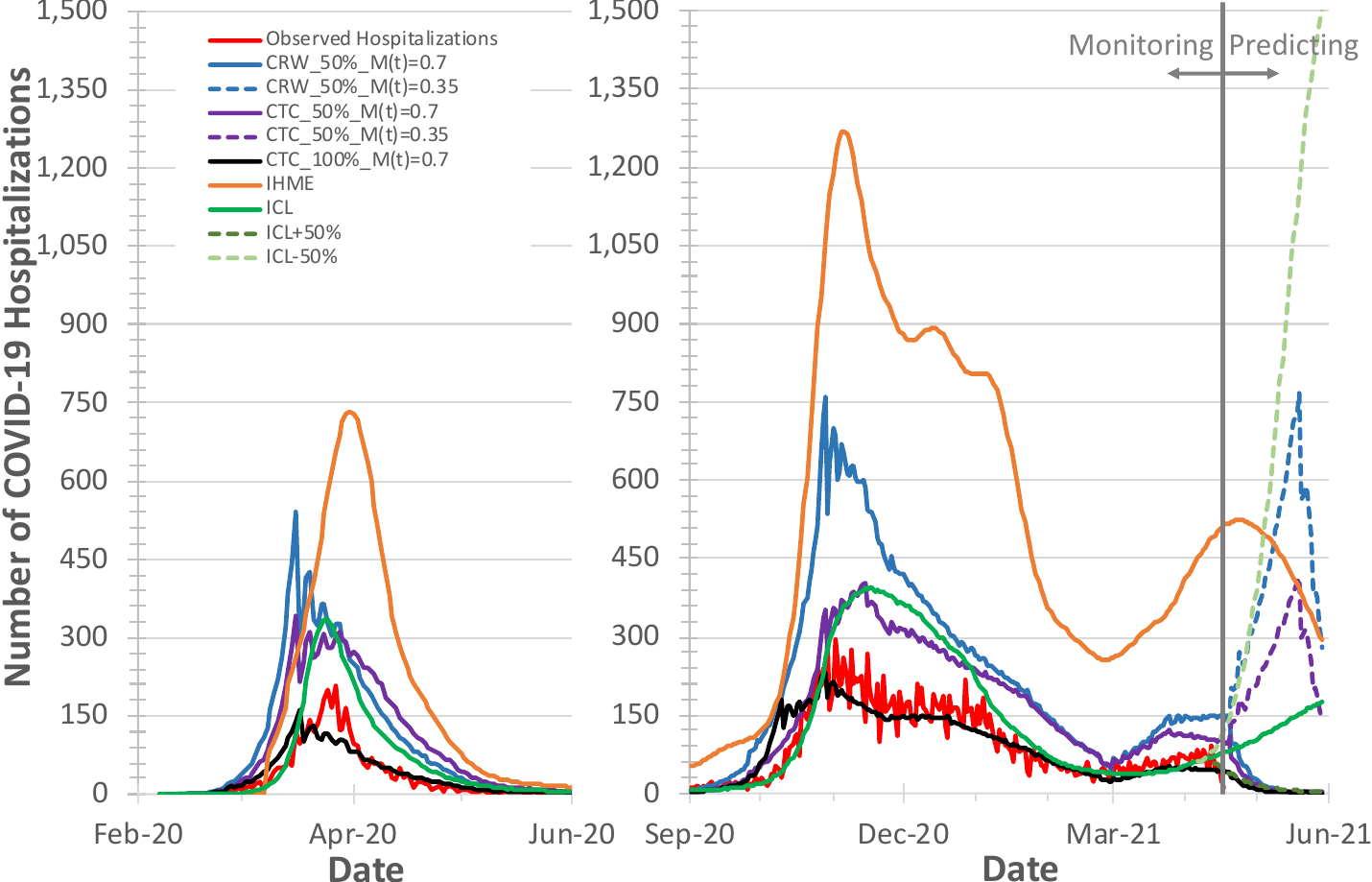}}
\caption{Observed and predicted number of COVID-19 hospitalizations.
We use two different environmental condition approaches, CRW and CTC with two certainty rate {levels} of 50\% and 100\%. 
We use two mitigation coefficient {values}, $M(t)$, of 0.35 and 0.7 for each configuration of our model during {19 April to 19 May 2021}.}
\label{FIG5}
\end{figure}

We evaluate COVIDHunter's {\emph{predicted}} daily number of COVID-19 {hospitalizations} in Figure~\ref{FIG5}.
We use the observed official number of hospitalizations as is.
Using the number of cases calculated with Equation~\ref{EQU2}, we find $X$ (hospitalizations-to-cases ratio) to be 4.288\% and 2.780\%, using CRW and CTC, respectively, during the second wave. 

We make five key observations based on Figure~\ref{FIG5}.
1) COVIDHunter (\texttt{CRW\_50\%\_M(t)=0.7}) with a certainty rate {level} of 50\% predicts on average $5.33\times$ {smaller} number of COVID-19 hospitalizations than that {calculated by the} IHME model.
2) {The} ICL model predicts the number of hospitalizations to be similar to that predicted by COVIDHunter (\texttt{CTC\_50\%\_M(t)=0.7}) during the first and the second waves.
{This suggests that both the ICL model and COVIDHunter (\texttt{CTC\_50\%\_M(t)=0.7}) consider that the actual number of COVID-19 hospitalizations is twice the observed number of COVID-19 hospitalizations}.
3) 
COVIDHunter with a certainty rate {level} of 100\% predicts the number of cases to perfectly fit the curve of the observed number of hospitalizations, reaching up to {231} hospitalized patients a day.
4) {Our model predicts that the number of COVID-19 hospitalizations reduces significantly (less than 5 daily hospitalized patients) within May 2021 if the mitigation measures that are applied nationwide in Switzerland are tightened ($M(t)$ increases from 0.55 to 0.7) for at least 30 days.
This is in line with what {the} ICL model ({\texttt{ICL+50\%}}) predicts, when ICL model is configured {to} strengthening the mitigation measures by 50\%.
If the authority decides to relax the mitigation measures to the lowest strength that has been applied during the year of 2020 ($M(t)$ drops from 0.55 to 0.35), then the daily expected number of hospitalizations \emph{exponentially} increases by an average of $5.1\times$ and $4.13\times$, becoming as high as the peak of the second wave (up to 767 daily hospitalized patients), using {the} CRW and CTC environmental approaches, respectively. 
ICL model predicts the situation to be worst, showing $2\times$ and $3.74\times$ higher number of hospitalizations than COVIDHunter \texttt{CRW\_50\%\_M(t)=0.35} and \texttt{CRW\_50\%\_M(t)=0.35}, respectively, when ICL model is configured {to} 50\% relaxation in the mitigation measures.
We provide a comprehensive evaluation for the effect of different mitigation coefficient values on the number of hospitalizations in the} \textcolor{blue}{Supplementary Materials, Section 2}. 
5) The use of {the} CTC approach for determining the environmental coefficient value yields a slightly different number {(on average $1.7\times$ less)} of hospitalizations compared to that provided by the use of {the} CRW approach.
This is expected as the CTC approach considers {only} the {monthly average} change in temperature, whereas the CRW approach considers the {daily} change in {\emph{several}} environmental conditions.

We conclude that {1) unlike the IBZ and LSHTM models, COVIDHunter is able to predict the number of hospitalizations and 2)} COVIDHunter provides more accurate estimation of the number of hospitalizations compared to that calculated by ICL (which provides {overestimation}) and IHME (which provides {late estimation}).
COVIDHunter predicts the number of COVID-19 hospitalizations in a {simple,} convenient and flexible way that requires calculating {only} the daily number of cases and the hospitalization-to-cases ratio, $C_{X}$. 

\subsection{Evaluating the {Predicted N}umber of COVID-19 Deaths}

\begin{figure}
\centerline{\includegraphics[width=0.6\linewidth]{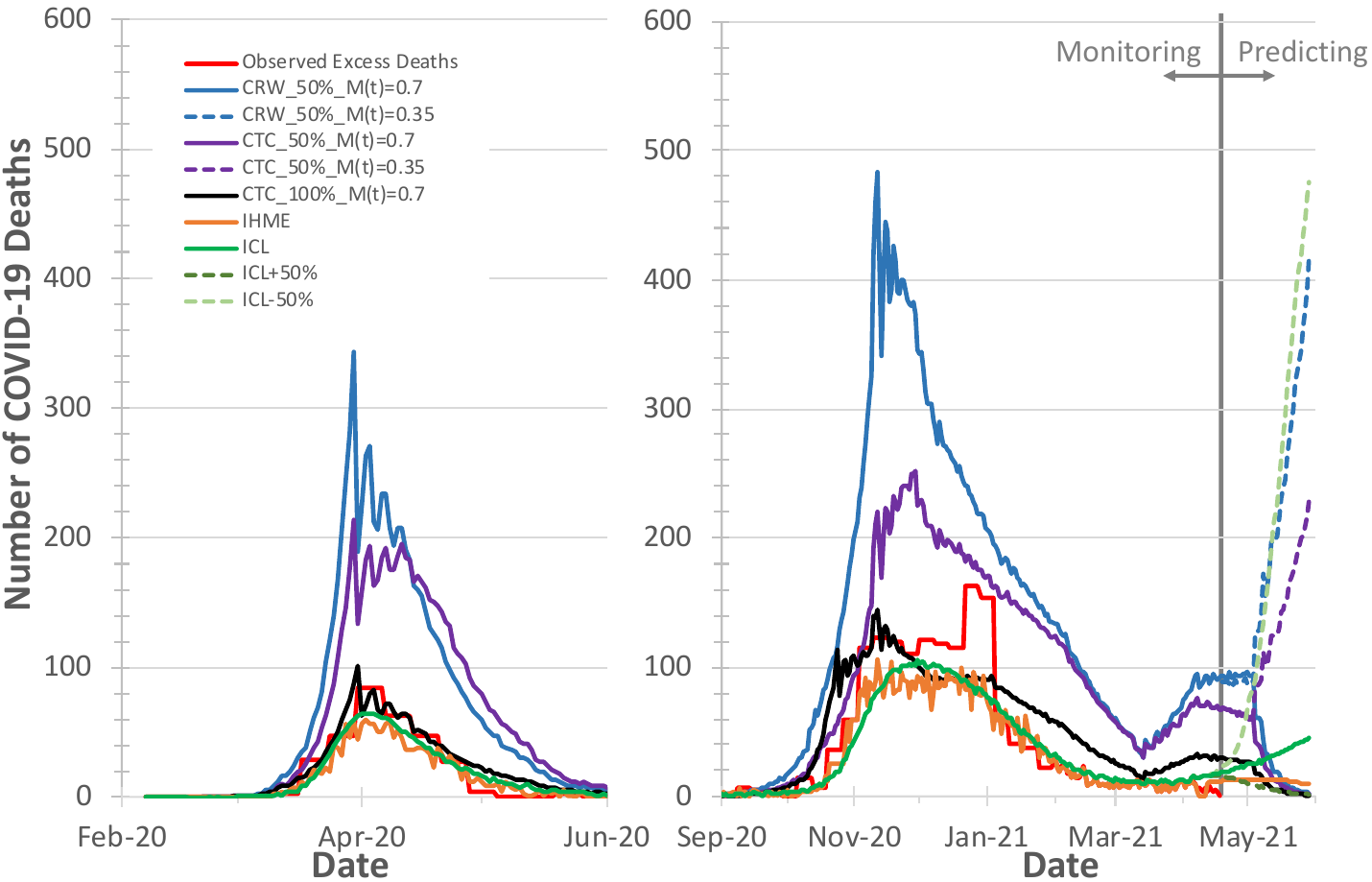}}
\caption{Observed and predicted number of COVID-19 deaths.
We use two different environmental condition approaches, CRW and CTC with two certainty rate levels of 50\% and 100\%. 
We use two mitigation coefficient values{, $M(t)$,} of 0.35 and 0.7 for each configuration of our model during {19 April to 19 May 2021}.}
\label{FIG6}
\end{figure}

We evaluate COVIDHunter's {\emph{predicted}} daily number of COVID-19 deaths in Figure~\ref{FIG6} after accounting for the 15-day shift (as we discuss in Section~\ref{results:modelvalidation}). 
We calculate the observed number of deaths as the number of {excess} deaths (Section~\ref{sec:HospDeath}) to account for uncertainty in reporting COVID-19 deaths. 
Using the number of cases calculated using Equation~\ref{EQU2}, we find $Y$ (deaths-to-cases ratio, using excess death data) to be 2.730\% and 1.739\%, using CRW and CTC, respectively, during the second wave.

We make three key observations based on Figure~\ref{FIG6}.
1) COVIDHunter with a certainty rate of 100\% predicts the number of deaths to perfectly fit the three curves of the observed number of \emph{excess} deaths, ICL deaths, and IHME deaths, {reaching up to 144 deaths} a day.
During the second wave, {the} ICL curve is shifted (late prediction) by 5-10 days from that of other models.
2) Similar to what we observe for the number of hospitalizations, our model predicts that the number of COVID-19 deaths significantly reduces {(reaching up to a single death a day)} with stricter mitigation measures ($M(t)$ increases from 0.55 to 0.7) maintained for at least the upcoming 30 days.
This is in line with what IHME model predicts.
Relaxing the mitigation measures ($M(t)$ drops from 0.55 to 0.35) \emph{exponentially} increases the death toll by {an average of $5.1\times$ and $4.13\times$, reaching up to 488} new daily deaths, as predicted by COVIDHunter using CRW and CTC environmental condition approaches, respectively.
COVIDHunter's prediction (\texttt{CRW\_50\%\_M(t)=0.35}) is in line with what ICL model predicts, when ICL model is configured as 50\% relaxation in the mitigation measures.
3) During the first wave, the use of a certainty rate of 50\% provides {$3\times$ and $2.7\times$ ($2.6\times$ and $1.7\times$ during the second wave)} higher number of deaths compared to that provided by ICL and IHME models, when COVIDHunter uses CRW and CTC environmental condition approaches, respectively.

\vspace{-10 pt}

{We conclude that 1) unlike {the} IBZ and LSHTM {models}, COVIDHunter is able to predict the number of deaths, 
2) COVIDHunter predicts the number of deaths to be similar to that predicted by {the} ICL and IHME models.
Yet, COVIDHunter provides more accurate estimation of other COVID-19 statistics ($R$, number of cases and hospitalizations) compared to ICL and IHME{,} as we comprehensively evaluate in the previous sections, and 3) COVIDHunter requires calculating {only} the daily number of cases and the deaths-to-cases ratio, $C_Y$}{, to predict the daily number of deaths}.

\section{{Summary} and Future Work} \label{sec:conclusion}
We demonstrate that we can monitor and predict the spread of COVID-19 in {an easy-to-use,} flexible, and validated way using our {new simulation} model, COVIDHunter.
We show how to flexibly configure our model for any scenario and easily extend it for different {mitigation} measures and {environmental} conditions.
{The use of} a small number of variables in our model {enables a simple and flexible yet powerful way of adapting} our model to different conditions for a {given} region. 
We demonstrate the importance of considering the effect of environmental changes on the spread of COVID-19 and how doing so can greatly improve simulation accuracy.
COVIDHunter flexibly offers the ability to directly make the best use of existing models that study the effect of one or both of environmental conditions and mitigation measures on the spread of COVID-19.

We benchmark our model against {major} alternative models of the COVID-19 pandemic that are used to assist governments.
{Compared to these models, COVIDHunter achieves more accurate estimation, provides no prediction delay, 
and provides ease of use {and high flexibility} due to the simple modeling approach} {that uses a small number of parameters}. 
{Using COVIDHunter, we demonstrate} that the spread of COVID-19 in Switzerland (as a case study) is still active {(i.e., $R$ > 1.0) and curbing this spread} requires maintaining the same or greater strength of the currently applied mitigation measures for at least another 30 days. 
{Using COVIDHunter on 19 April 2021, we predicted that the strength of the mitigation measures applied nationwide in Switzerland was relaxed by 14-21\% on 1 March 2021.
The predicted drop is in line with the increase in the observed official number of cases, hospitalizations, and deaths by 16\%, 3.9\%, and 2.8\%, respectively, (as shown by the Federal Office of Public Health in Switzerland} \url{www.covid19.admin.ch}).
{Relaxing the mitigation measures further by 30-40\% on 19 April for 30 days would exponentially increase the number of cases, hospitalizations, and deaths by $5.1\times$}. 
We provide insights on the effect of each change in the strength of the applied mitigation measure on the number of daily cases, hospitalizations, and deaths.
We make all the data, statistical analyses, and a well-documented model implementation publicly and freely available to enable full reproducibility and {help} society and decision-makers to {accurately and openly} review the current situation {and estimate future impact of decisions}.

We suggest {and plan at least} five main directions/additions to further improve the {predictive} power {and benefits} of our {COVIDHunter} model.
1) Clustering the population based on age-groups.
This has potential different effects on, for example, population, environmental conditions, mitigation {measures}~\citep{bhatia2020estimating, bi2020epidemiology}.
2) Considering vaccinated persons as another new category of persons in a population.
3) Considering reinfection after {immunity}~\citep{lumley2020antibody}.
4) Considering the average number of households (or population density){, as well as other potential population-level effects,} while calculating the number of new infected persons caused by an infected person.
5) Considering different strains of the COVID-19 virus by allowing for multiple base reproduction numbers.
{Our goal is to update COVIDHunter with such improvements and capabilities while keeping {its} simplicity, ease of use, and flexibility of {its} modeling strategy}.
\vspace{-10 pt}


\bibliographystyle{natbib}

\bibliography{document}

\begin{thebibliography}{}

\bibitem[Alser {\em et~al.}(2021)Alser, Kim, Alserr, Tell, and
  Mutlu]{alser2021covidhunterv1}
Alser, M., Kim, J.~S., Alserr, N.~A., {\em et~al.} (2021).
\newblock Covidhunter: An accurate, flexible, and environment-aware open-source
  covid-19 outbreak simulation model.
\newblock {\em arXiv preprint arXiv:2102.03667v1\/}.

\bibitem[Anastassopoulou {\em et~al.}(2020)Anastassopoulou, Russo, Tsakris, and
  Siettos]{anastassopoulou2020data}
Anastassopoulou, C., Russo, L., Tsakris, A., and Siettos, C. (2020).
\newblock {Data-based analysis, modelling and forecasting of the COVID-19
  outbreak}.
\newblock {\em PloS one\/}, {\bf 15}(3), e0230405.

\bibitem[Andersen {\em et~al.}(2020)Andersen, Rambaut, Lipkin, Holmes, and
  Garry]{andersen2020proximal}
Andersen, K.~G., Rambaut, A., Lipkin, W.~I., {\em et~al.} (2020).
\newblock {The proximal origin of SARS-CoV-2}.
\newblock {\em Nature medicine\/}, {\bf 26}(4), 450--452.

\bibitem[Ashcroft {\em et~al.}(2020)Ashcroft, Huisman, Lehtinen, Bouman,
  Althaus, Regoes, and Bonhoeffer]{ashcroft2020covid}
Ashcroft, P., Huisman, J.~S., Lehtinen, S., {\em et~al.} (2020).
\newblock {COVID-19 infectivity profile correction}.
\newblock {\em Swiss Medical Weekly\/}, {\bf 150}(3132).

\bibitem[Bhatia and Klausner(2020)Bhatia and Klausner]{bhatia2020estimating}
Bhatia, R. and Klausner, J. (2020).
\newblock {Estimating individual risks of COVID-19-associated hospitalization
  and death using publicly available data}.
\newblock {\em PloS one\/}, {\bf 15}(12), e0243026.

\bibitem[Bi {\em et~al.}(2020)Bi, Wu, Mei, Ye, Zou, Zhang, Liu, Wei, Truelove,
  Zhang, {\em et~al.}]{bi2020epidemiology}
Bi, Q., Wu, Y., Mei, S., {\em et~al.} (2020).
\newblock {Epidemiology and transmission of COVID-19 in 391 cases and 1286 of
  their close contacts in Shenzhen, China: a retrospective cohort study}.
\newblock {\em The Lancet Infectious Diseases\/}, {\bf 20}(8), 911--919.

\bibitem[Brauner {\em et~al.}(2020)Brauner, Mindermann, Sharma, Johnston,
  Salvatier, Gaven{\v{c}}iak, Stephenson, Leech, Altman, Mikulik, {\em
  et~al.}]{brauner2020inferring}
Brauner, J.~M., Mindermann, S., Sharma, M., {\em et~al.} (2020).
\newblock {Inferring the effectiveness of government interventions against
  COVID-19}.
\newblock {\em Science\/}.

\bibitem[Chang {\em et~al.}(2020)Chang, Harding, Zachreson, Cliff, and
  Prokopenko]{chang2020modelling}
Chang, S.~L., Harding, N., Zachreson, C., {\em et~al.} (2020).
\newblock {Modelling transmission and control of the COVID-19 pandemic in
  Australia}.
\newblock {\em Nature Communications\/}.

\bibitem[Davies {\em et~al.}(2020)Davies, Kucharski, Eggo, Gimma, Edmunds,
  Jombart, O'Reilly, Endo, Hellewell, Nightingale, {\em
  et~al.}]{davies2020effects}
Davies, N.~G., Kucharski, A.~J., Eggo, R.~M., {\em et~al.} (2020).
\newblock {Effects of non-pharmaceutical interventions on COVID-19 cases,
  deaths, and demand for hospital services in the UK: a modelling study}.
\newblock {\em The Lancet Public Health\/}.

\bibitem[de~Souza {\em et~al.}(2020)de~Souza, Buss, da~Silva~Candido, Carrera,
  Li, Zarebski, Vincenti-Gonzalez, Messina, da~Silva~Sales, dos Santos~Andrade,
  {\em et~al.}]{de2020epidemiological}
de~Souza, W.~M., Buss, L.~F., da~Silva~Candido, D., {\em et~al.} (2020).
\newblock {Epidemiological and clinical characteristics of the early phase of
  the COVID-19 epidemic in Brazil}.
\newblock {\em Nature Human Behaviour\/}, {\bf 4}, 856–865.

\bibitem[Du~Toit(2020)Du~Toit]{du2020outbreak}
Du~Toit, A. (2020).
\newblock {Outbreak of a novel coronavirus}.
\newblock {\em Nature Reviews Microbiology\/}, {\bf 18}(3), 123--123.

\bibitem[Fares(2013)Fares]{fares2013factors}
Fares, A. (2013).
\newblock {Factors influencing the seasonal patterns of infectious diseases}.
\newblock {\em International journal of preventive medicine\/}, {\bf 4}(2),
  128.

\bibitem[Ferguson {\em et~al.}(2020)Ferguson, Laydon, Nedjati-Gilani, Imai,
  Ainslie, Baguelin, Bhatia, Boonyasiri, Cucunub{\'a}, Cuomo-Dannenburg, {\em
  et~al.}]{ferguson2020report}
Ferguson, N., Laydon, D., Nedjati-Gilani, G., {\em et~al.} (2020).
\newblock {Report 9: Impact of non-pharmaceutical interventions (NPIs) to
  reduce COVID19 mortality and healthcare demand}.
\newblock {\em Imperial College London\/}, {\bf 10}, 77482.

\bibitem[Fisman(2012)Fisman]{fisman2012seasonality}
Fisman, D. (2012).
\newblock {Seasonality of viral infections: mechanisms and unknowns}.
\newblock {\em Clinical Microbiology and Infection\/}, {\bf 18}(10), 946--954.

\bibitem[Flaxman {\em et~al.}(2020)Flaxman, Mishra, Gandy, Unwin, Mellan,
  Coupland, Whittaker, Zhu, Berah, Eaton, {\em et~al.}]{flaxman2020estimating}
Flaxman, S., Mishra, S., Gandy, A., {\em et~al.} (2020).
\newblock {Estimating the effects of non-pharmaceutical interventions on
  COVID-19 in Europe}.
\newblock {\em Nature\/}, {\bf 584}(7820), 257--261.

\bibitem[Hale {\em et~al.}(2020)Hale, Petherick, Phillips, and
  Webster]{hale2020variation}
Hale, T., Petherick, A., Phillips, T., and Webster, S. (2020).
\newblock {Variation in government responses to COVID-19}.
\newblock {\em Blavatnik school of government working paper\/}, {\bf 31}.

\bibitem[Hilton and Keeling(2020)Hilton and Keeling]{Hilton2020Rnumber}
Hilton, J. and Keeling, M.~J. (2020).
\newblock {Estimation of country-level basic reproductive ratios for novel
  Coronavirus (SARS-CoV-2/COVID-19) using synthetic contact matrices}.
\newblock {\em PLOS Computational Biology\/}, {\bf 16}(7), 1--10.

\bibitem[Huisman {\em et~al.}(2020)Huisman, Scire, Angst, Neher, Bonhoeffer,
  and Stadler]{huisman2020estimation}
Huisman, J.~S., Scire, J., Angst, D.~C., {\em et~al.} (2020).
\newblock {Estimation and worldwide monitoring of the effective reproductive
  number of SARS-CoV-2}.
\newblock {\em medrxiv\/}.

\bibitem[Ioannidis {\em et~al.}(2020)Ioannidis, Cripps, and
  Tanner]{ioannidis2020forecasting}
Ioannidis, J.~P., Cripps, S., and Tanner, M.~A. (2020).
\newblock {Forecasting for COVID-19 has failed}.
\newblock {\em International journal of forecasting\/}.

\bibitem[Jagannathan and Wang(2021)Jagannathan and
  Wang]{jagannathan2021immunity}
Jagannathan, P. and Wang, T.~T. (2021).
\newblock {Immunity after SARS-CoV-2 infections}.
\newblock {\em Nature Immunology\/}, pages 1--2.

\bibitem[Kampf {\em et~al.}(2020)Kampf, Todt, Pfaender, and
  Steinmann]{kampf2020persistence}
Kampf, G., Todt, D., Pfaender, S., and Steinmann, E. (2020).
\newblock {Persistence of coronaviruses on inanimate surfaces and their
  inactivation with biocidal agents}.
\newblock {\em Journal of Hospital Infection\/}, {\bf 104}(3), 246--251.

\bibitem[Kobayashi {\em et~al.}(2020)Kobayashi, Jung, Linton, Kinoshita,
  Hayashi, Miyama, Anzai, Yang, Yuan, Akhmetzhanov, {\em
  et~al.}]{kobayashi2020communicating}
Kobayashi, T., Jung, S.-m., Linton, N.~M., {\em et~al.} (2020).
\newblock {Communicating the Risk of Death from Novel Coronavirus Disease
  (COVID-19)}.
\newblock {\em Journal of Clinical Medicine\/}, {\bf 9}(2).

\bibitem[Lauer {\em et~al.}(2020)Lauer, Grantz, Bi, Jones, Zheng, Meredith,
  Azman, Reich, and Lessler]{lauer2020incubation}
Lauer, S.~A., Grantz, K.~H., Bi, Q., {\em et~al.} (2020).
\newblock {The incubation period of coronavirus disease 2019 (COVID-19) from
  publicly reported confirmed cases: estimation and application}.
\newblock {\em Annals of internal medicine\/}, {\bf 172}(9), 577--582.

\bibitem[Li {\em et~al.}(2020)Li, Guan, Wu, Wang, Zhou, Tong, Ren, Leung, Lau,
  Wong, {\em et~al.}]{li2020early}
Li, Q., Guan, X., Wu, P., {\em et~al.} (2020).
\newblock {Early transmission dynamics in Wuhan, China, of novel
  coronavirus--infected pneumonia}.
\newblock {\em New England Journal of Medicine\/}.

\bibitem[Lumley {\em et~al.}(2020)Lumley, O’Donnell, Stoesser, Matthews,
  Howarth, Hatch, Marsden, Cox, James, Warren, {\em
  et~al.}]{lumley2020antibody}
Lumley, S.~F., O’Donnell, D., Stoesser, N.~E., {\em et~al.} (2020).
\newblock {Antibody status and incidence of SARS-CoV-2 infection in health care
  workers}.
\newblock {\em New England Journal of Medicine\/}.

\bibitem[Moriyama {\em et~al.}(2020)Moriyama, Hugentobler, and
  Iwasaki]{moriyama2020seasonality}
Moriyama, M., Hugentobler, W.~J., and Iwasaki, A. (2020).
\newblock {Seasonality of respiratory viral infections}.
\newblock {\em Annual review of virology\/}, {\bf 7}.

\bibitem[Noll {\em et~al.}(2020)Noll, Aksamentov, Druelle, Badenhorst, Ronzani,
  Jefferies, Albert, and Neher]{noll2020covid}
Noll, N.~B., Aksamentov, I., Druelle, V., {\em et~al.} (2020).
\newblock {COVID-19 Scenarios: an interactive tool to explore the spread and
  associated morbidity and mortality of SARS-CoV-2}.
\newblock {\em MedRxiv\/}.

\bibitem[Pachetti {\em et~al.}(2020)Pachetti, Marini, Benedetti, Giudici,
  Mauro, Storici, Masciovecchio, Angeletti, Ciccozzi, Gallo, {\em
  et~al.}]{pachetti2020emerging}
Pachetti, M., Marini, B., Benedetti, F., {\em et~al.} (2020).
\newblock {Emerging SARS-CoV-2 mutation hot spots include a novel
  RNA-dependent-RNA polymerase variant}.
\newblock {\em Journal of Translational Medicine\/}, {\bf 18}, 1--9.

\bibitem[Petropoulos and Makridakis(2020)Petropoulos and
  Makridakis]{petropoulos2020forecasting}
Petropoulos, F. and Makridakis, S. (2020).
\newblock {Forecasting the novel coronavirus COVID-19}.
\newblock {\em PloS one\/}, {\bf 15}(3), e0231236.

\bibitem[Phan(2020)Phan]{phan2020genetic}
Phan, T. (2020).
\newblock {Genetic diversity and evolution of SARS-CoV-2}.
\newblock {\em Infection, genetics and evolution\/}, {\bf 81}, 104260.

\bibitem[Prata {\em et~al.}(2020)Prata, Rodrigues, and
  Bermejo]{prata2020temperature}
Prata, D.~N., Rodrigues, W., and Bermejo, P.~H. (2020).
\newblock {Temperature significantly changes COVID-19 transmission in (sub)
  tropical cities of Brazil}.
\newblock {\em Science of the Total Environment\/}, page 138862.

\bibitem[Rahman {\em et~al.}(2020)Rahman, Sadraddin, and
  Porreca]{rahman2020basic}
Rahman, B., Sadraddin, E., and Porreca, A. (2020).
\newblock {The basic reproduction number of SARS-CoV-2 in Wuhan is about to die
  out, how about the rest of the World?}
\newblock {\em Reviews in Medical Virology\/}, page e2111.

\bibitem[Reichmuth {\em et~al.}(2021)Reichmuth, Hodcroft, Riou, Althaus,
  Schibler, Eckerle, Kaiser, Huber, Trkola, Hasse, Nilsson, Buonomano, and
  Neher]{Reichmuth2021}
Reichmuth, M., Hodcroft, E., Riou, J., {\em et~al.} (2021).
\newblock {Transmission of SARS-CoV-2 variants in Switzerland}.
\newblock \url{https://ispmbern.github.io/covid-19/variants/index.pdf}.
\newblock Accessed: 2021-1-30.

\bibitem[Reiner {\em et~al.}(2020)Reiner, Barber, Collins, Zheng, Adolph,
  Albright, Antony, Aravkin, Bachmeier, Bang-Jensen, {\em
  et~al.}]{reiner2020modeling}
Reiner, R.~C., Barber, R.~M., Collins, J.~K., {\em et~al.} (2020).
\newblock {Modeling COVID-19 scenarios for the United States}.
\newblock {\em Nature Medicine\/}.

\bibitem[Riddell {\em et~al.}(2020)Riddell, Goldie, Hill, Eagles, and
  Drew]{riddell2020effect}
Riddell, S., Goldie, S., Hill, A., {\em et~al.} (2020).
\newblock {The effect of temperature on persistence of SARS-CoV-2 on common
  surfaces}.
\newblock {\em Virology journal\/}, {\bf 17}(1), 1--7.

\bibitem[Russell {\em et~al.}(2020)Russell, Golding, Hellewell, Abbott, Wright,
  Pearson, van Zandvoort, Jarvis, Gibbs, Liu, {\em
  et~al.}]{russell2020reconstructing}
Russell, T.~W., Golding, N., Hellewell, J., {\em et~al.} (2020).
\newblock {Reconstructing the early global dynamics of under-ascertained
  COVID-19 cases and infections}.
\newblock {\em BMC medicine\/}, {\bf 18}(1), 1--9.

\bibitem[Shi {\em et~al.}(2020)Shi, Hu, Peng, Tang, Wang, Su, Luo, Wu, Zhang,
  Zhang, {\em et~al.}]{shi2020effective}
Shi, Q., Hu, Y., Peng, B., {\em et~al.} (2020).
\newblock {Effective control of SARS-CoV-2 transmission in Wanzhou, China}.
\newblock {\em Nature medicine\/}, pages 1--8.

\bibitem[Slifka and Gao(2020)Slifka and Gao]{slifka2020presymptomatic}
Slifka, M.~K. and Gao, L. (2020).
\newblock {Is presymptomatic spread a major contributor to COVID-19
  transmission?}
\newblock {\em Nature Medicine\/}, {\bf 26}(10), 1531--1533.

\bibitem[Toyoshima {\em et~al.}(2020)Toyoshima, Nemoto, Matsumoto, Nakamura,
  and Kiyotani]{toyoshima2020sars}
Toyoshima, Y., Nemoto, K., Matsumoto, S., {\em et~al.} (2020).
\newblock {SARS-CoV-2 genomic variations associated with mortality rate of
  COVID-19}.
\newblock {\em Journal of human genetics\/}, {\bf 65}(12), 1075--1082.

\bibitem[Tradigo {\em et~al.}(2020)Tradigo, Guzzi, Kahveci, and
  Veltri]{tradigo2020method}
Tradigo, G., Guzzi, P.~H., Kahveci, T., and Veltri, P. (2020).
\newblock {A method to assess COVID-19 infected numbers in Italy during peak
  pandemic period}.
\newblock In {\em 2020 IEEE International Conference on Bioinformatics and
  Biomedicine (BIBM)\/}, pages 3017--3020. IEEE.

\bibitem[Wang {\em et~al.}(2020)Wang, Tang, Feng, and Lv]{wang2020high}
Wang, J., Tang, K., Feng, K., and Lv, W. (2020).
\newblock {High temperature and high humidity reduce the transmission of
  COVID-19}.
\newblock {\em Available at SSRN 3551767\/}.

\bibitem[Wei {\em et~al.}(2020)Wei, Li, Chiew, Yong, Toh, and
  Lee]{wei2020presymptomatic}
Wei, W.~E., Li, Z., Chiew, C.~J., {\em et~al.} (2020).
\newblock {Presymptomatic Transmission of SARS-CoV-2—Singapore, January
  23--March 16, 2020}.
\newblock {\em Morbidity and Mortality Weekly Report\/}, {\bf 69}(14), 411.

\bibitem[Wilke and Bergstrom(2020)Wilke and Bergstrom]{wilke2020predicting}
Wilke, C.~O. and Bergstrom, C.~T. (2020).
\newblock {Predicting an epidemic trajectory is difficult}.
\newblock {\em Proceedings of the National Academy of Sciences\/}, {\bf
  117}(46), 28549--28551.

\bibitem[Xie and Zhu(2020)Xie and Zhu]{xie2020association}
Xie, J. and Zhu, Y. (2020).
\newblock {Association between ambient temperature and COVID-19 infection in
  122 cities from China}.
\newblock {\em Science of the Total Environment\/}, {\bf 724}, 138201.

\bibitem[Xu {\em et~al.}(2020)Xu, Rahmandad, Gupta, DiGennaro, Ghaffarzadegan,
  Amini, and Jalali]{xu2020modest}
Xu, R., Rahmandad, H., Gupta, M., {\em et~al.} (2020).
\newblock {The Modest Impact of Weather and Air Pollution on COVID-19
  Transmission}.
\newblock {\em medRxiv\/}.

\end{thebibliography}

\includepdf[pages=-]{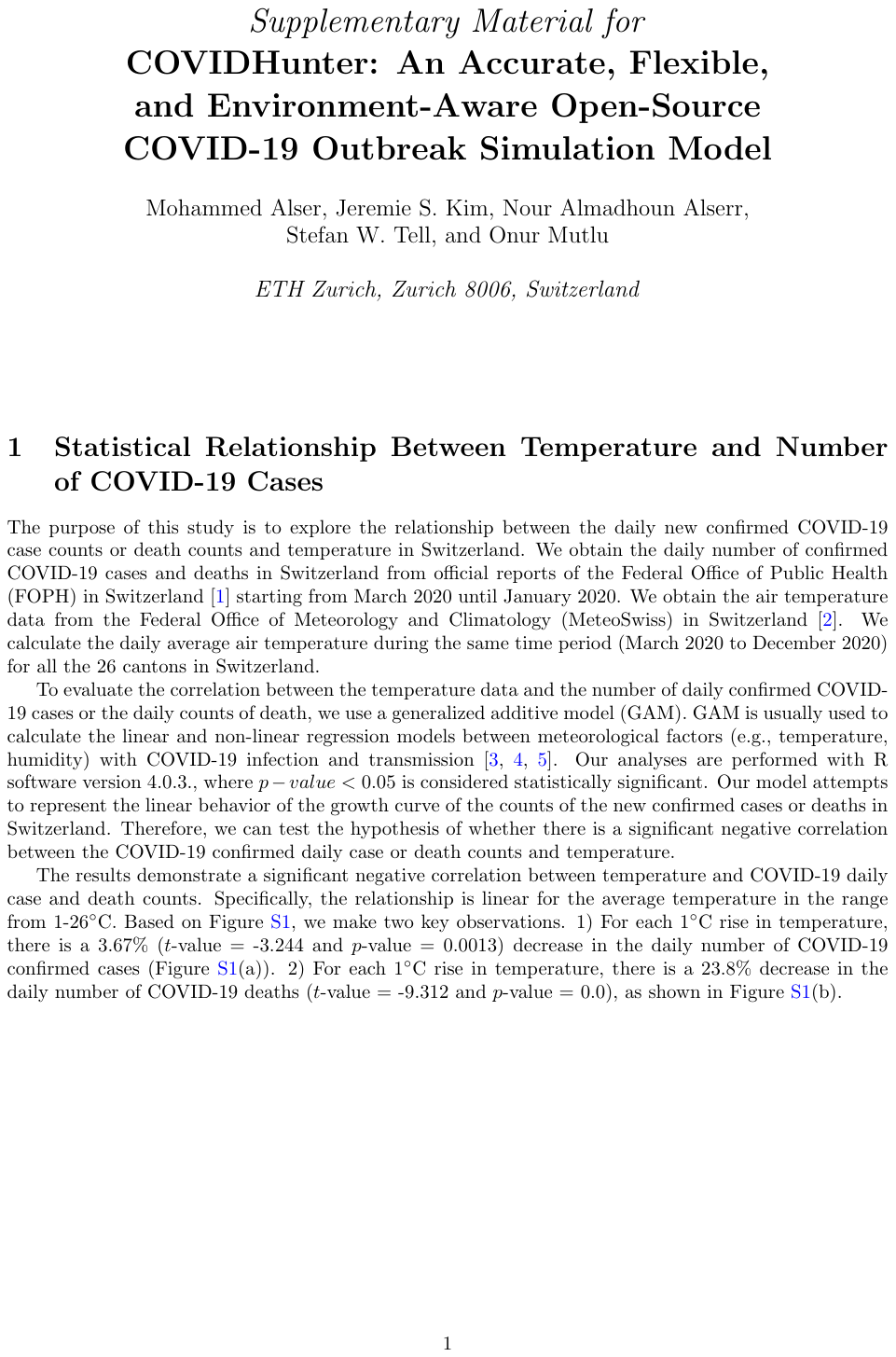}

\end{document}


\maketitle

\section{Statistical Relationship Between Temperature and Number of COVID-19 Cases}
\label{SuppCases}
The purpose of this study is to explore the {relationship} between the daily new confirmed COVID-19 case counts or death counts and temperature in Switzerland.
We obtain the daily number of confirmed COVID-19 cases and deaths in Switzerland from official reports of the {Federal Office of Public Health (FOPH)} in Switzerland~\cite{coronaCH} starting from March 2020 until January 2020. We obtain the {air} temperature data from the Federal Office of Meteorology and Climatology (MeteoSwiss) in Switzerland~\cite{forecastCH}. We calculate the daily average {air} temperature during the same time period (March 2020 to December 2020) for all the 26 cantons in Switzerland.

To evaluate the {correlation} between the temperature data and the number of daily confirmed COVID-19 cases or the daily counts of death, we use a generalized additive model (GAM). 
GAM is usually used to calculate the linear and non-linear regression models between meteorological factors {(e.g., temperature, humidity)} with COVID-19 infection and transmission~\cite{liu2020impact,prata2020temperature,xie2020association}.
{Our} analyses are performed with R software version 4.0.3., where {$p-value<0.05$} is considered statistically significant. 
Our model attempts to represent the linear behavior of the growth curve of the counts of the new confirmed cases or deaths in Switzerland. Therefore, {we can test the hypothesis of} whether there is a {significant} negative correlation between the COVID-19 confirmed daily case or death counts and temperature. 

\begin{figure}
\centering
\subfigure[]{
\includegraphics[width=0.5\textwidth]{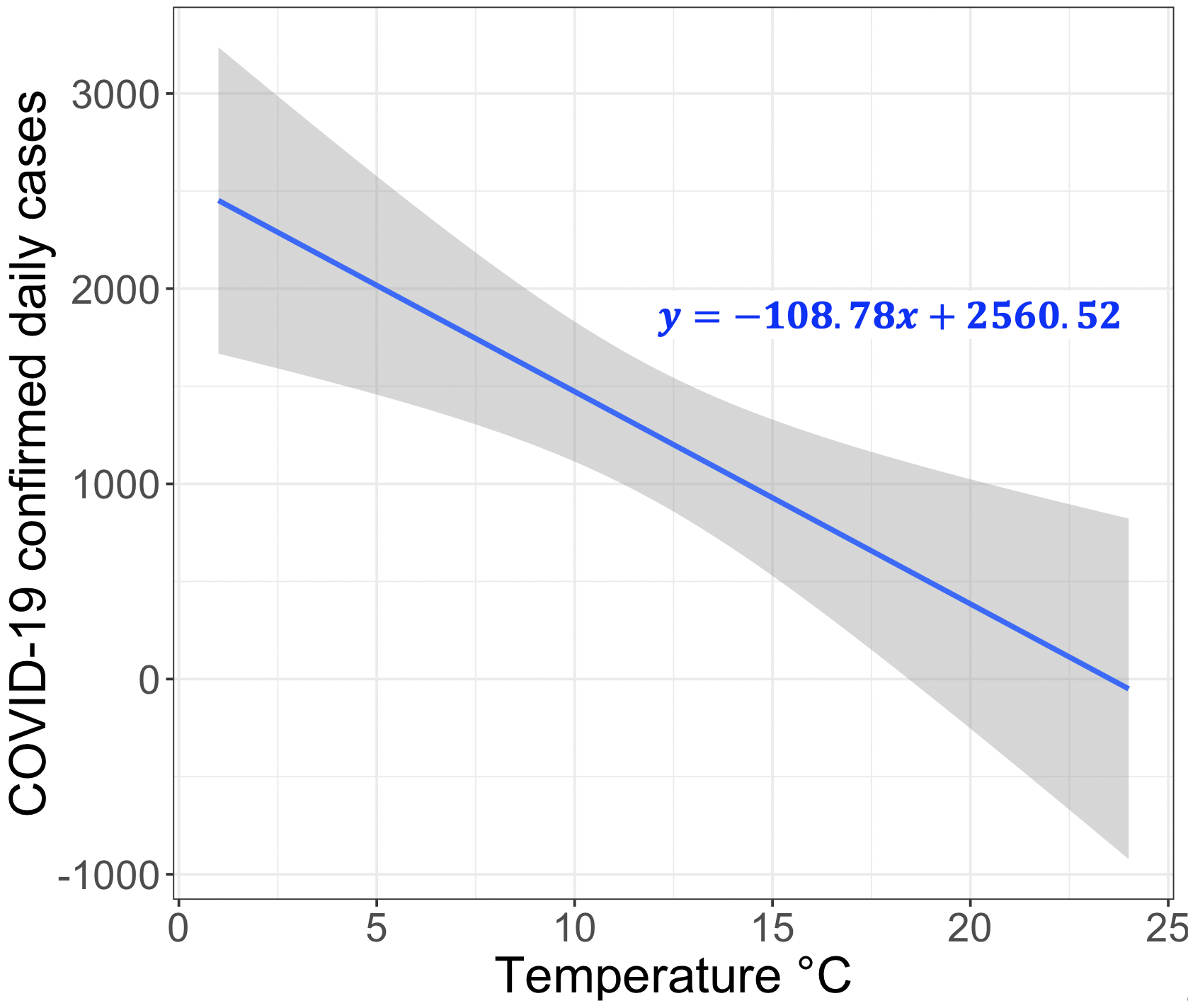}}%
\qquad
\subfigure[]{
\includegraphics[width=0.5\textwidth]{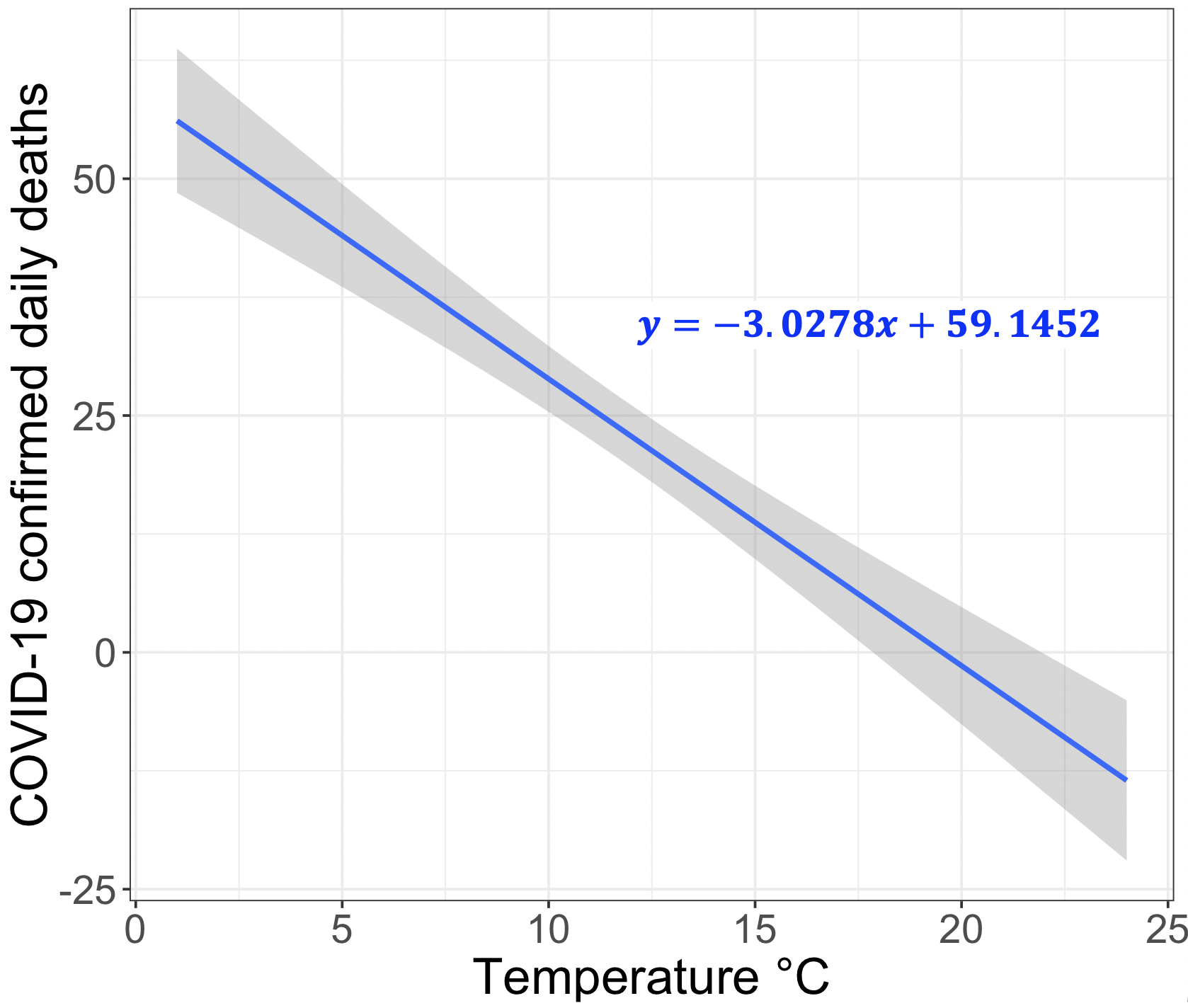}}%
\caption{Correlation between temperature and COVID-19 confirmed (a) case count and (b) death count in 26 cantons of Switzerland.} 
\label{fig:01}
\end{figure}

The results {demonstrate} a {significant} negative {correlation} between temperature and COVID-19 daily case and death counts. Specifically, the relationship {is} linear for the average temperature in the range from 1-26$^{\circ}$C. 
Based on Figure~\ref{fig:01}, {we make two key observations. 1) For each 1$^{\circ}$C rise in temperature, there is a 3.67\% ($t$-value = −3.244 and $p$-value = 0.0013) decrease in the daily number of COVID-19 confirmed cases} (Figure~\ref{fig:01}(a)).
2) {For each 1$^{\circ}$C rise in temperature, there is a 23.8\% decrease in the daily number of COVID-19 deaths} ($t$-value = −9.312 and $p$-value = 0.0), as shown in Figure~\ref{fig:01}(b).


\newpage

\section{Evaluating the Effect of Different Mitigation Coefficient Values on COVIDHunter's Predicted Number of Cases, Hospitalizations, and Deaths}

\begin{figure}[H]
\centerline{\includegraphics[width=0.85\linewidth]{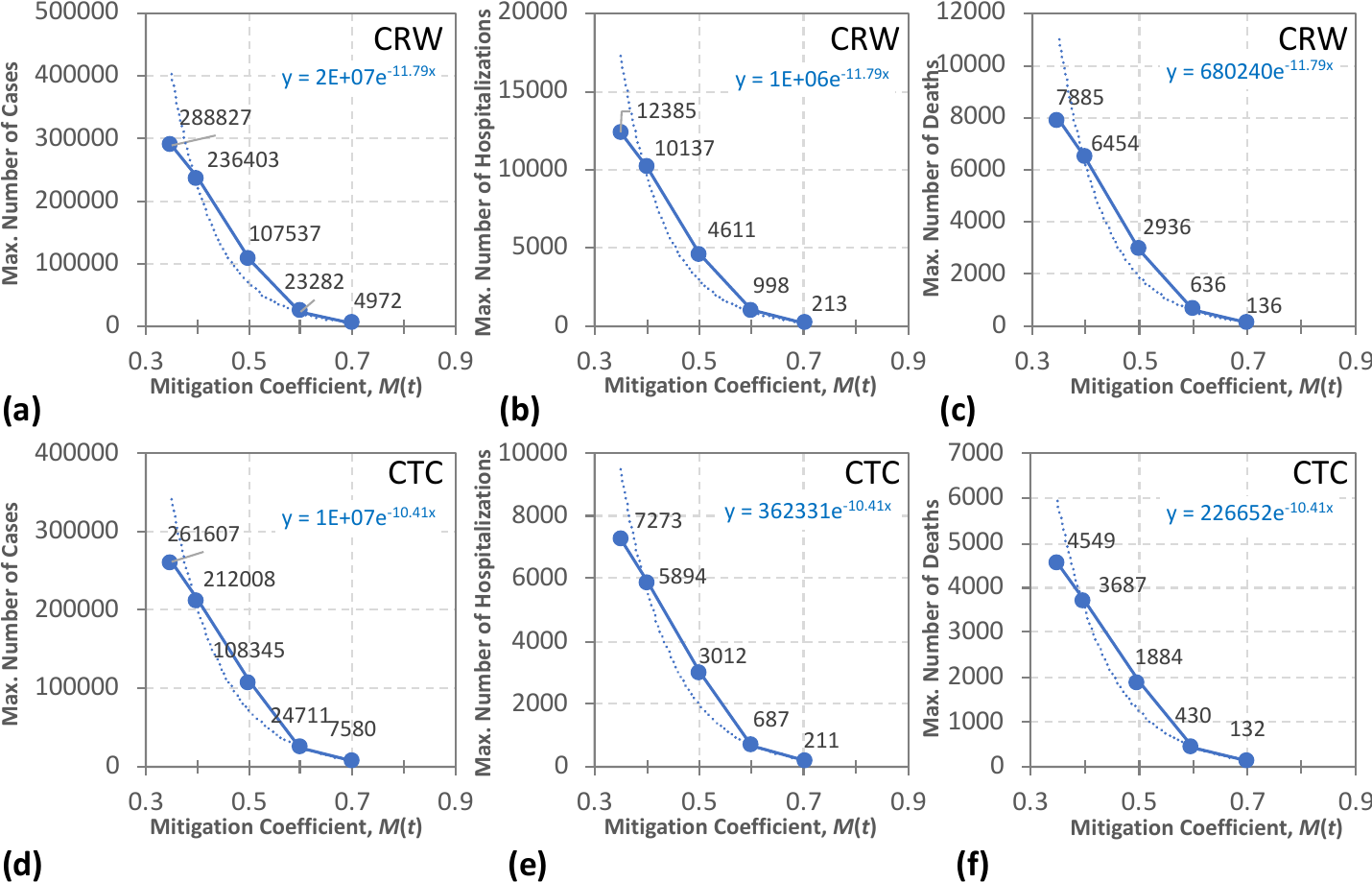}}
\caption{The maximum {of daily} number of COVID-19 cases, hospitalizations, and deaths as predicted by COVIDHunter {over 22 January to 31 March 2021}.
We use five mitigation coefficient, $M(t)$, values of 0.35, 0.4, 0.5, 0.6, and 0.7 for each configuration of our model during {22 January to 22 February 2021}.
We use two different environmental condition approaches, CRW \textbf{(a)-(c)} and CTC \textbf{(d)-(f)} with a certainty rate level of 50\%.
Dashed line represents exponential model fit to data.
}
\label{FIG8}
\end{figure}

\begin{figure}[H]
\centerline{\includegraphics[width=0.85\linewidth]{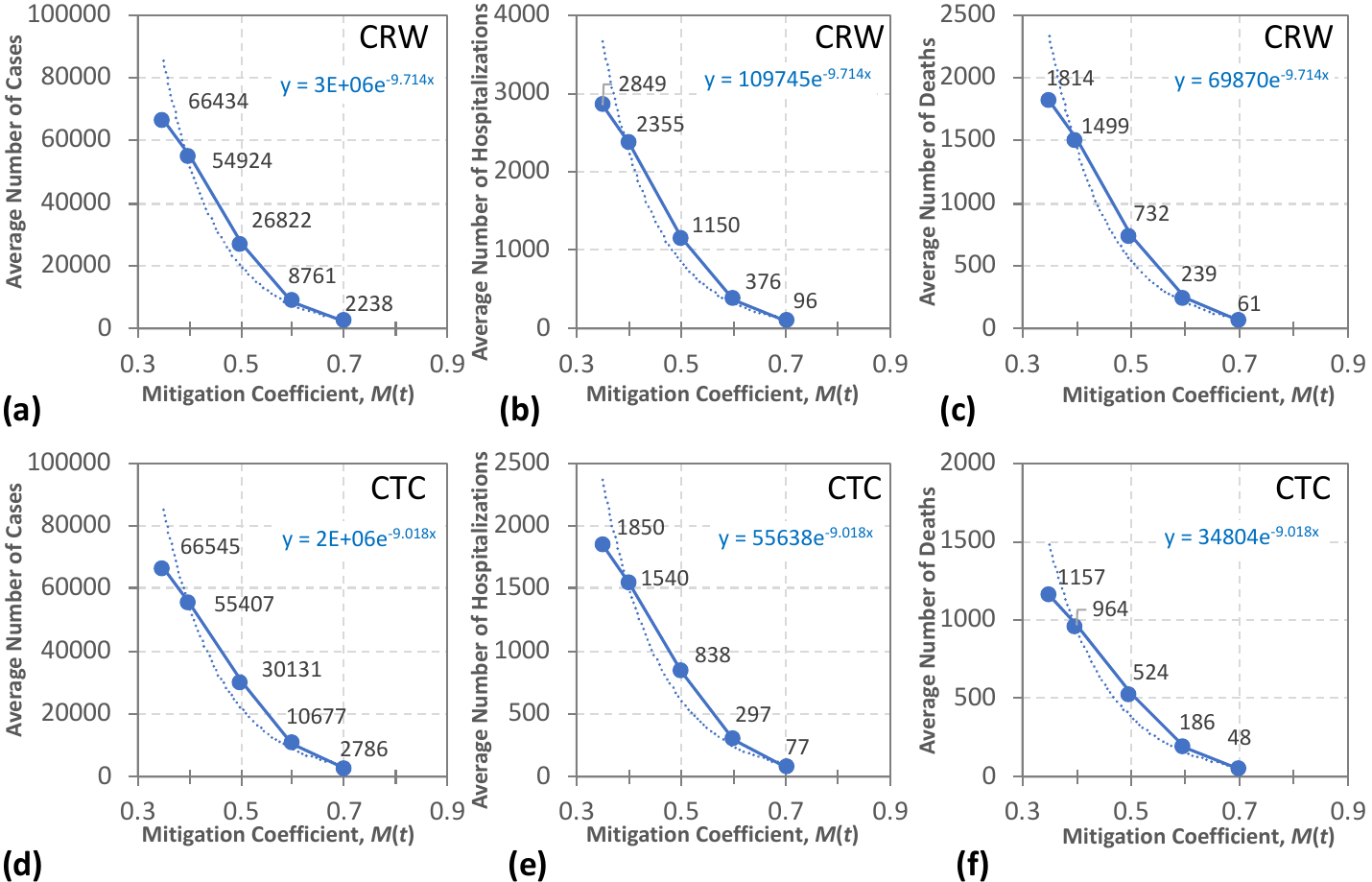}}
\caption{The average {of daily} number of COVID-19 cases, hospitalizations, and deaths as predicted by COVIDHunter {over 22 January to 31 March 2021}.
We use five mitigation coefficient, $M(t)$, values of 0.35, 0.4, 0.5, 0.6, and 0.7 for each configuration of our model during {22 January to 22 February 2021}.
We use two different environmental condition approaches, CRW \textbf{(a)-(c)} and CTC \textbf{(d)-(f)} with a certainty rate level of 50\%.
Dashed line represents exponential model fit to data.}
\label{FIG9}
\end{figure}

Using COVIDHunter, we predict the number of COVID-19 cases, hospitalizations, and deaths during {22 January to 31 March 2021}.
We show the maximum and the average of daily number of COVID-19 cases, hospitalizations, and deaths {over 22 January to 31 March 2021} in Figures~\ref{FIG8} and ~\ref{FIG9}, respectively.
We use two environmental condition approaches, CRW and CTC, with a certainty rate level of 50\%.
We assume five mitigation coefficient, $M(t)$, values of 0.35, 0.4, 0.5, 0.6, and 0.7 for each configuration of COVIDHunter during {22 January to 22 February 2021}.

This range of mitigation coefficient values covers the lowest (i.e., $M(t)$=0.35) and the highest (i.e., $M(t)$=0.7) strengths of mitigation measures that have been applied during the year of 2020.

Based on Figures~\ref{FIG8} and ~\ref{FIG9}, we make three key observations.
1) COVIDHunter predicts that the \emph{maximum} {of daily} number of COVID-19 cases, hospitalizations, and deaths {over 22 January to 31 March 2021} would be 4972, 213, and 136, respectively, using CRW and $M(t)$=0.7, as we show in Figure~\ref{FIG8}(a-c). 
Using our environmental condition approach, CTC, and $M(t)$=0.7, the \emph{maximum} {of daily} number of COVID-19 cases {over 22 January to 31 March 2021} would be 7580 and the \emph{maximum} {of daily} number of COVID-19 hospitalizations and deaths would be almost same as that calculated by COVIDHunter with CRW, as we show in Figure~\ref{FIG8}(d-f). 
2) Relaxing the mitigation measures by 50\% ($M$ is changed from 0.7 to 0.35) \emph{exponentially} increases the \emph{maximum} {of daily} number of cases, hospitalizations, and deaths by $58\times$, reaching up to 288827, 12385, and 7885, respectively, as predicted by COVIDHunter with the CRW approach (Figure~\ref{FIG8}(a-c)).
Using the CTC appraoch and $M(t)$=0.35, COVIDHunter predicts an \emph{exponential} increase in the \emph{maximum} {of daily} number of cases, hospitalizations, and deaths by \emph{only} $34.5\times$, as we show in Figure~\ref{FIG8}(a-c).
This is expected as the CTC approach considers only the drop in temperature rather than the average effect of many environmental conditions as the CRW approach does.
3) Relaxing the mitigation measures by 50\% ($M$ is changed from 0.7 to 0.35) causes the \emph{daily} number of cases, hospitalizations, and deaths to \emph{exponentially} increase by an average of $29.6\times$ and $23.8\times$ {over 22 January to 31 March 2021} using CRW and CTC environmental approaches, respectively, as we show in Figure~\ref{FIG9}.

We conclude that COVIDHunter provides {flexible} evaluation of the effect of different strength of the past and current mitigation measures on the {number of COVID-19 cases, hospitalizations, and deaths}.
COVIDHunter evaluates the applied mitigation measures with high flexibility of configuring the environmental coefficient and mitigation coefficient, which helps society and {decision-makers} to accurately review the current situation and estimate future impact of decisions.

\section{Evaluated Datasets}
Our experimental evaluation uses a large number of different real datasets, including 1) daily $R$ number values, 2) observed daily number of COVID-19 cases, 3) observed daily number of COVID-19 hospitalizations, 4) observed daily number of COVID-19 deaths, 5) number of excess deaths, 6) the estimated strength of mitigation measures as calculated by the Oxford Stringency Index, 7) estimation of COVID-19 statistics as calculated by existing state-of-the-art simulation models, ICL, IHME, LSHTM, and IBZ, from seven different sources as we list below. The raw datasets are provided in the \textcolor{red}{GitHub page of COVIDHunter and in the Supplementary {Excel File}.}

\begin{itemize}
\item Observed COVID-19 statistics (R number values and number of cases, hospitalizations, and deaths)
\begin{itemize}
\item Official reports (January 7, 2021): \url{https://www.covid19.admin.ch/en/overview}
\item Smoothed data (January 7, 2021): \url{https://ourworldindata.org/coronavirus/country/switzerland?country=~CHE}
\end{itemize}

\item Excess deaths:
\begin{itemize}
\item Information: \url{https://www.bfs.admin.ch/bfs/en/home/statistics/health/state-health/mortality-causes-death.html}
\item Direct link (January 7, 2021): \url{https://www.bfs.admin.ch/bfs/en/home/statistics/health/state-health/mortality-causes-death.assetdetail.12607335.html}
\end{itemize}

\item Oxford Stringency Index
\begin{itemize}
\item \url{https://www.bsg.ox.ac.uk/research/research-projects/coronavirus-government-response-tracker#data}
\end{itemize}

\item Imperial College London (ICL) Model:
\begin{itemize}
\item Information: \url{https://mrc-ide.github.io/global-lmic-reports/}
\item Direct link (January 15, 2021): \url{https://github.com/mrc-ide/global-lmic-reports/raw/master/data/2021-01-30_v7.csv.zip}
\end{itemize}

\item Institute for Health Metrics and Evaluation (IHME) Model:
\begin{itemize}
\item Information: \url{https://mrc-ide.github.io/global-lmic-reports/}
\item Direct link (January 15, 2021): \url{http://www.healthdata.org/covid/data-downloads}
\end{itemize}

\item The London School of Hygiene & Tropical Medicine (LSHTM) Model:
\begin{itemize}
\item Information: \url{https://cmmid.github.io/topics/covid19/global_cfr_estimates.html}
\item Direct link (January 15, 2021): \url{https://raw.githubusercontent.com/cmmid/cmmid.github.io/master/topics/covid19/reports/under_reporting_estimates/under_ascertainment_estimates.csv}
\end{itemize}

\item The Theoretical Biology Group at ETH Zurich (IBZ) Model:
\begin{itemize}
\item Information: \url{https://ibz-shiny.ethz.ch/covid-19-re-international/}
\item Direct link (January 15, 2021): \url{https://github.com/covid-19-Re/dailyRe-Data}    
\end{itemize}
\end{itemize}

\bibliographystyle{unsrt}
\bibliography{document}